\newif\ifshortver
\shortverfalse
\newif\ifasymm
\asymmfalse

\ifshortver
\documentclass[conference]{IEEEtran}
\else
\documentclass[english,onecolumn]{IEEEtran}
\fi

\IEEEoverridecommandlockouts

\usepackage[T1]{fontenc}
\usepackage[utf8]{inputenc}
\usepackage{color}
\usepackage[cmex10]{amsmath}
\usepackage{amsthm}
\DeclareMathOperator{\arctanh}{arctanh}
\usepackage{amssymb}
\usepackage{esint}
\usepackage{pgfplots}
\usepackage{pgfplotstable}
\usepackage{soul}
\usepackage{bbm}

\makeatletter
\theoremstyle{plain}
\newtheorem{thm}{\protect\theoremname}
\theoremstyle{plain}

\theoremstyle{plain}
\newtheorem{lem}{\protect\lemmaname}
\theoremstyle{definition}

\theoremstyle{remark}
\newtheorem{rem}{\protect\remarkname}

\usepackage{mathrsfs}
\usepackage{algorithm,algpseudocode}

\usepackage{colortbl}
\definecolor{lightgray}{rgb}{0.9,0.9,0.9}
\definecolor{lightred}{rgb}{1,0.8,0.8}
\definecolor{lightgreen}{rgb}{0.6,1,0.6}
\definecolor{lightyellow}{rgb}{1,1,0.5}
\definecolor{lightgrey}{rgb}{0.8,0.8,0.8}

\allowdisplaybreaks

\makeatother

\usepackage{babel}
\providecommand{\corollaryname}{Corollary}
\providecommand{\lemmaname}{Lemma}
\providecommand{\theoremname}{Theorem}
\providecommand{\remarkname}{Remark}
\pgfplotsset{compat=1.17}
\begin{document}
\title{Weighted Parity-Check Codes for Channels with State and Asymmetric Channels}
\author{%
Chih Wei Ling, Yanxiao Liu, \textit{Student Member, IEEE}, and Cheuk Ting Li, \textit{Member, IEEE}
\thanks{
The work of Cheuk Ting Li was supported in part by the Hong Kong Research Grant Council Grant ECS No. CUHK 24205621, and the Direct Grant for Research, The Chinese University of Hong Kong (Project ID: 4055133).

This paper was presented in part at the 2022 IEEE International Symposium on Information Theory (ISIT).

Chih Wei Ling, Yanxiao Liu and Cheuk Ting Li are with the Department of Information Engineering, The Chinese University of Hong
Kong, Hong Kong SAR of China. Email: chihweiLing@link.cuhk.edu.hk, yanxiaoliu@link.cuhk.edu.hk,  ctli@ie.cuhk.edu.hk. 
Chih Wei Ling and Yanxiao Liu contribute equally to this paper.
}
}
\maketitle
\begin{abstract}
\ifshortver
\else
In this paper, we introduce a new class of codes, called weighted parity-check codes, where each parity-check bit has a weight that indicates its likelihood to be one (instead of fixing each parity-check bit to be zero). It is applicable to a wide range of settings, e.g. asymmetric channels, channels with state and/or cost constraints, and the Wyner-Ziv problem, and can provably achieve the capacity. For the channels with state (Gelfand-Pinsker) setting, the proposed coding scheme has two advantages compared to the nested linear code. First, it achieves the capacity of any channel with state (e.g. asymmetric channels). Second, simulation results show that the proposed code achieves a smaller error rate compared to the nested linear code. We also discuss a sparse construction where the belief propagation algorithm can be applied to improve the coding efficiency.
\fi
\end{abstract}


\medskip{}

\section{Introduction}

In conventional (linear or nonlinear) code construction, the codebook is a set, and whether a bit sequence belongs to the codebook is a binary choice.
In the recent work on code construction by the Poisson functional representation~\cite{sfrl_trans,li2021unified}, the codebook is instead a ``fuzzy set'', where each bit sequence has a weight that corresponds to the likelihood that the sequence is selected. It was shown in~\cite{li2021unified} that this ``weighted codebook'' construction has several advantages (e.g. better theoretical guarantee for channels with state, which will be explained later). Nevertheless, the random weight assignment in~\cite{li2021unified} is unstructured, and does not allow efficient encoding and decoding algorithms.

In this paper, which is the complete version of~\cite{ling2022weighted},\footnote{The conference paper \cite{ling2022weighted} includes the description (but not the complete proof) of the weighted parity-check codes for channels with state \cite{gelfand1980coding} and experiments with dense parity-check matrices. Compared to \cite{ling2022weighted}, this complete version also includes the complete analysis on the weighted parity-check codes for channels with state and the Wyner-Ziv problem~\cite{wyner1976ratedistort}, and the belief propagation coding algorithms for sparse parity-check matrices.} we present a general code construction based on this weighted codebook idea, but with a linear structure similar to conventional linear codes. Instead of fixing each parity-check bit to zero, we assign a weight to each parity-check bit that indicates its likelihood to be one. We call this \emph{weighted parity-check (WPC) codes}. We will discuss its applications to
channels with state and asymmetric channels.

In a channel with state~\cite{shannon1958channels,Kuznetsov1974Tsybakov,gelfand1980coding,el2011network}, the channel statistics depend on a state variable, which is not fully known and varies over the transmission.
We review some works related to channels with state information available at the encoder non-causally (i.e., the Gelfand-Pinsker setting \cite{gelfand1980coding}). 
The problem of memory with stuck-at foaults was studied by Kuznetsov and Tsybakov \cite{Kuznetsov1974Tsybakov}, where multicoding/subcodebooks is used to establish the capacity.
Their result was generalized by Gelfand and Pinsker \cite{gelfand1980coding} and Heegard and El Gamal \cite{Heegard1980} to the discrete memoryless channel (DMC) with discrete memoryless (DM) state, where the capacity was characterized.
Costa~\cite{Costa1983} proved the capacity theorem for the Gaussian channel with additive Gaussian state (i.e., "writing on dirty paper").
Refer to~\cite{verdu2012nonasymp,yassaee2013oneshot,watanabe2015nonasymp} for finite-blocklength analysis of the Gelfand-Pinsker setting.

Note that the coding schemes used in the direct part of the proofs of the aforementioned works are based on unstructured codes which are impractical.
To realize a structured code construction, Zamir \textit{et al.} \cite{Zamir2002Nested} and Barron \textit{et al.} \cite{barron2003duality}  considered the nested linear codes for binary-Hamming information embedding~\cite{chen2001quantization,swanson1998multimedia}, and nested lattice codes for Gaussian-quadratic information embedding. 
Nested linear codes were studied previously by Wyner~\cite{wyner1974recent} for the Slepian-Wolf problem~\cite{slepian1973noiseless}, and by Shamai \textit{et al.}~\cite{shamai1998systematic} and Pradhan and Ramchandran~\cite{pradhan2003distributed} for the Wyner-Ziv problem~\cite{wyner1976ratedistort}. The connection between codes for the Wyner-Ziv problem and codes for channels with state was observed in~\cite{chou1999duality,Zamir2002Nested,barron2003duality}.
In \cite{martinian2006low2},  Martinian and Wainwright used a class of sparse graphical codes to generate practical nested linear codes.
Nested constructions using polar codes were studied in~\cite{korada2010polar}.

The aforementioned structured codes are for symmetric channels. For structured codes for the general (symmetric/asymmetric) channel with state, Padakandla and Pradhan~\cite{padakandla2011nested} utilized nested linear codes together with joint typicality encoding and decoding to construct coding schemes for channels with state and broadcast channels.
Efficient coding scheme based on non-linear polar codes was proposed in~\cite{gad2016asymmetric}. 
Ghaddar \textit{et al.}~\cite{ghaddar2021lego} proposed a method for constructing coding schemes for both the asymmetric channels and the Gelfand-Pinsker problem by leveraging existing codes for symmetric point-to-point channels.

All the aforementioned works use some variants of the binning/subcodebook construction, where each message is associated with a subcodebook of input sequences (either randomly generated, or as a coset of a linear code in~\cite{Zamir2002Nested,barron2003duality}), and the encoder chooses an input sequence within the subcodebook close to the state sequence. The weighted codebook construction~\cite{li2021unified} eliminates the need of subcodebooks, and gives finite-blocklength and second-order error bounds sharper than previous finite-blocklength results in~\cite{verdu2012nonasymp,yassaee2013oneshot,watanabe2015nonasymp}.
An intuitive reason is that the subcodebook construction forces the encoder to choose a sequence within the subcodebook (a non-fuzzy set), which may result in having a sequence far from the state sequence. In the weighted codebook construction, we have a fuzzy set instead, so the encoder can trade-off between choosing a high-weight sequence (higher likelihood of being chosen by the decoder) and a low-weight sequence (which may be closer to the state sequence) depending on the state. The downside is that the weighted codebook in~\cite{li2021unified} is unstructured.

The weighted parity-check codes in this paper combine the advantages of the linear construction~\cite{Zamir2002Nested,barron2003duality} and the weighted codebook~\cite{li2021unified}, giving a structured code construction that provably achieves the capacity, and has good practical performance.
Experiment results show that our codes attain a smaller error rate compared to nested linear codes~\cite{Zamir2002Nested,barron2003duality}.
Moreover, our codes can be applied on a sparse parity-check matrix in a manner similar to the low-density parity-check codes~\cite{gallager1962low,mackay1999good}, which allows the use of the belief propagation algorithm \cite{Kschischang2001, Loeliger2004,richardson_urbanke_2008} to improve the coding efficiency.

Another application of our codes is to asymmetric channels (e.g. Z-channel), which arises in storage technologies such as flash memories~\cite{ahlswede2002unidirectional, klove2011systematic}. It was observed in~\cite{ghaddar2021lego} that a code for binary asymmetric channels can be obtained from a code for symmetric channels with state by setting the state sequence to zero. Our code can be applied to general (symmetric/asymmetric) channels with or without state, which is more general than~\cite{ghaddar2021lego} which applies only to the case where the distribution of the channel input conditional on the state is a binary symmetric channel. Compared to~\cite{padakandla2011nested} which is also general, our scheme does not require joint typicality encoding and decoding. Instead, it admits an encoding and decoding scheme with a simple product structure where belief propagation can be applicable.

\subsection{Other Related Works}

Fuzzy codes are generalizations of conventional codes where the codebook is a fuzzy set~\cite{von1982fuzzy,hall1990fuzzy,tsafack2018fuzzy}. The code in this paper can also be regarded as a fuzzy code. Nevertheless, this paper focuses on a concrete capacity-achieving coding scheme, whereas~\cite{von1982fuzzy,hall1990fuzzy,tsafack2018fuzzy} are more about properties (e.g. distances) of general fuzzy codes in the context of coding theory, and have not discussed capacity-achieving properties. We also remark that the application of soft sets (generalization of fuzzy sets) to coding was studied in~\cite{ali2018new}. 
The construction used in this paper is unrelated to~\cite{von1982fuzzy,hall1990fuzzy,tsafack2018fuzzy,ali2018new}.

In decoding algorithms for linear codes for soft-information channels (e.g. AWGN), weighted parity-check information is sometimes utilized (e.g. \cite{kou2001low}), where the weights
come from the soft-information channel instead of being associated with the
parity-check bit itself (as in this paper). In the analysis on Gallager codes in \cite{tanaka2002information}, parity-check bits were assumed to be observed with error for the sake of analytical tractability, though this soft parity-check assumption was only used in the analysis of decoding error instead of the code construction.

\subsection*{Notations}

Logarithm and entropy are to the base $2$.  Natural logarithm is
written as $\ln(x)$. The binary cross entropy function is 
\begin{equation}
H_{b}(\phi,\psi):=-\phi\log\psi-(1-\phi)\log(1-\psi),\label{eq:hb}
\end{equation}
and the binary entropy function is $H_{b}(\phi):=H_{b}(\phi,\phi)$.
The binary symmetric channel with crossover probability $\beta$ is written as $\mathrm{BSC}(\beta)$.
We write $\mathbf{1}^{n}=[1,\ldots,1]\in\mathbb{R}^{n}$ and $\mathbf{0}^{n}=[0,\ldots,0]$. 
\ifshortver
\else
For a distribution $P_X$ over $\mathcal{X}$, the typical set is $\mathcal{T}_{\epsilon}^{(n)}(P_{X}):=\{ \mathbf{x} \in \mathcal{X}^n:\,\forall x \in \mathcal{X} .\,|\hat{P}_{\mathbf{x}}(x)-P_{X}(x)|\le\epsilon P_{X}(x)\}$, where $\hat{P}_{\mathbf{x}}(x)$ is the empirical distribution of $\mathbf{x}$. The conditional typical set is $\mathcal{T}_{\epsilon}^{(n)}(P_{Y|X}|\mathbf{x}):=\{ \mathbf{y} \in \mathcal{Y}^n:\, (\mathbf{x}, \mathbf{y}) \in \mathcal{T}_{\epsilon}^{(n)}(P_{X,Y})\}$.
The finite field of order $2$ is denoted as $\mathbb{F}_{2}$. Addition between finite field vectors in $\mathbb{F}_{2}^n$ is denoted as ``$\mathbf{x} \oplus \mathbf{y}$''.
\fi

\section{Weighted Parity-Check codes}

Consider the channel coding setting where the encoder encodes the
message $\mathbf{m}\in\mathbb{F}_{2}^{k}$ into the codeword $\mathbf{x}\in\mathbb{F}_{2}^{n}$.
The decoder receives $\mathbf{y}\in\mathbb{F}_{2}^{n}$ (a noise-corrupted
version of $\mathbf{x}$) and recovers the message as $\hat{\mathbf{m}}\in\mathbb{F}_{2}^{k}$.
We now describe the construction of the weighted parity-check codes. Let $\mathbf{H}\in\mathbb{F}_{2}^{n\times n}$ be a full-rank
matrix, called the \emph{full
parity-check matrix}. We assume that $\mathbf{H}$ is a uniformly randomly
chosen matrix among the set of $n\times n$ full-rank matrices with entries in $\mathbb{F}_{2}$, though other constructions (e.g. random
sparse matrix) are also possible (see Section~\ref{sec:ldpc}). For a \emph{bias vector} $\mathbf{q}=[q_{1},\ldots,q_{n}]\in[0,1]^{n}$,
define the $\mathbf{q}$\emph{-weight} of a vector $\mathbf{u}\in\mathbb{F}_{2}^{n}$
as
\begin{align*}
w_{\mathbf{q}}(\mathbf{u}) & :=\prod_{i=1}^{n}q_{i}^{u_{i}}(1-q_{i})^{1-u_{i}}  =2^{-\sum_{i=1}^{n}H_{b}(u_{i},q_{i})}.
\end{align*}
Intuitively, $w_{\mathbf{q}}(\mathbf{u})$ is the probability of $\mathbf{u}$
assuming the entries $u_{i}\sim\mathrm{Bern}(q_{i})$ are independent
across $i$.

Given the bias vectors $\mathbf{p},\mathbf{q}\in[0,1]^{n}$ (we call
$\mathbf{p}$ the \emph{codeword bias}, and $\mathbf{q}$ the
\emph{parity bias}), the \emph{query function} is given by
\begin{equation}
f_{\mathbf{H}}(\mathbf{p},\mathbf{q}):=\mathrm{argmax}_{\mathbf{x}\in\mathbb{F}_{2}^{n}}\,w_{\mathbf{p}}(\mathbf{x})w_{\mathbf{q}}(\mathbf{x}\mathbf{H}^{T}).\label{eq:fH}
\end{equation}

The encoder has two parameters: the \emph{encoder codeword bias function}
$\mathbf{p}_{e}:\mathbb{F}_{2}^{k}\to[0,1]^{n}$ which maps the message
$\mathbf{m}\in\mathbb{F}_{2}^{k}$ (and other information available
at the encoder) to a bias vector $\mathbf{p}_{e}(\mathbf{m})$, and
the \emph{encoder parity bias function} $\mathbf{q}_{e}:\mathbb{F}_{2}^{k}\to[0,1]^{n}$.
The actual encoding function is
\[
\mathbf{m}\mapsto\mathbf{x}=f_{\mathbf{H}}\left(\mathbf{p}_{e}(\mathbf{m}),\,\mathbf{q}_{e}(\mathbf{m})\right).
\]

The decoder likewise has two parameters: the \emph{decoder codeword
and parity bias functions} $\mathbf{p}_{d},\mathbf{q}_{d}:\mathbb{F}_{2}^{n}\to[0,1]^{n}$.
The decoding function is
\begin{equation}
\mathbf{y}\mapsto\hat{\mathbf{m}}=\left[(\hat{\mathbf{x}}\mathbf{H}^{T})_{1},\ldots,\,(\hat{\mathbf{x}}\mathbf{H}^{T})_{k}\right],\label{eq:decoder}
\end{equation}
where
\[
\hat{\mathbf{x}}:=f_{\mathbf{H}}\left(\mathbf{p}_{d}(\mathbf{y}),\,\mathbf{q}_{d}(\mathbf{y})\right).
\]
Note that we use the first $k$ bits of $\mathbf{x}\mathbf{H}^{T}$ to represent the message, and the remaining bits for parity-check bits, hence have the decoding function in~\eqref{eq:decoder}.

One advantage of the product form in the query function $f_{\mathbf{H}}$
is that it allows the use of belief propagation in the encoding and
decoding function (if $\mathbf{H}$ is sparse; see Section~\ref{sec:ldpc}). The bits in $\mathbf{x}\mathbf{H}^{T}$ (after the first $k$ bits)
can be regarded as ``soft parity-check bits'' that are only observed
with noise. Loosely speaking, the encoder's prior distributions of
the parity-check bits are $\mathbf{P}((\mathbf{x}\mathbf{H}^{T})_{i}=1)=(\mathbf{q}_{e}(\mathbf{m}))_{i}$,
and the decoder's posterior distributions are $\mathbf{P}((\mathbf{x}\mathbf{H}^{T})_{i}=1)=(\mathbf{q}_{d}(\mathbf{y}))_{i}$.
When $(\mathbf{q}_{e}(\mathbf{m}))_{i} = 0$ (or $1$), the parity-check bit $(\mathbf{x}\mathbf{H}^{T})_{i}$ is fixed to $0$ (or $1$). When $(\mathbf{q}_{e}(\mathbf{m}))_{i} = 1/2$, the parity-check bit is unused (equally likely to be $0$ or $1$).

The definition of the weighted parity-check codes is quite general. To recover the conventional linear code, we take
\begin{align*}
\mathbf{p}_{e}(\mathbf{m}) & =\frac{1}{2}\mathbf{1}^{n}  = \Big[\frac{1}{2},\ldots, \frac{1}{2}\Big], &\mathbf{q}_{e}(\mathbf{m}) & =[\mathbf{m},\,\mathbf{0}^{n-k}],\\
\mathbf{p}_{d}(\mathbf{y}) & =\beta\mathbf{1}^{n}+(1-2\beta)\mathbf{y}, &\mathbf{q}_{d}(\mathbf{y}) & =\Big[\frac{1}{2}\mathbf{1}^{k},\,  \mathbf{0}^{n-k}\Big],
\end{align*}
where we assume the channel $\mathbf{x}\to\mathbf{y}$ is $\mathrm{BSC}(\beta)$.
Note that $w_{\mathbf{p}_{d}(\mathbf{y})}(\mathbf{x})=P(\mathbf{x}|\mathbf{y})$
is the posterior distribution of $\mathbf{x}$.

To apply this construction to asymmetric channels, we may change $\mathbf{p}_{e}(\mathbf{m})$ to $\alpha\mathbf{1}^{n}$, $\alpha \in (0,1)$ to introduce bias to the bits in $\mathbf{x}$. Nevertheless, note that since each entry of $\mathbf{q}_{e}(\mathbf{m})$ is $0$ or $1$, we know that $\mathbf{x} = [\mathbf{m},\,\mathbf{0}^{n-k}]\mathbf{H}^{-T}$ is fixed by $\mathbf{m}$ and does not depend on $\alpha$. To allow a biased distribution of $\mathbf{x}$, we have to ``soften'' the parity-check bits so that $\mathbf{x}$ does not only depend on $\mathbf{m}$. This will be discussed in the next section as a special case of Theorem~\ref{thm:capacity}.

\begin{rem}\label{rem:finitefield}
We remark that the weighted parity-check codes can be extended naturally
to any finite field $\mathbb{F}_{l}$ of order $l>2$. In that case,
each entry of $\mathbf{p}_{e},\mathbf{q}_{e},\mathbf{p}_{d},\mathbf{q}_{d}$
would be a distribution over $\mathbb{F}_{l}$ (i.e., a vector in
the probability simplex over $\mathbb{F}_{l}$) instead of a number
in $[0,1]$, and $w_{\mathbf{q}}(\mathbf{u})$ would be the probability of $\mathbf{u}$
assuming $u_{i}$ follows $q_i$ (a distribution over $\mathbb{F}_{l}$) and are independent
across $i$.
\end{rem}

\section{Channels with State\label{sec:state}}

Consider the setting where the channel has a state that is available
noncausally to the encoder \cite{gelfand1980coding}. The state sequence
$\mathbf{s}=[s_{1},\ldots,s_{n}]$, where $s_{i}\in\mathcal{S}$ (not
necessarily binary), $s_{i}\stackrel{iid}{\sim}P_{S}$, is available
at the encoder. Given $\mathbf{s}$, the encoder encodes the message $\mathbf{m}\in\mathbb{F}_{2}^{k}$
into $\mathbf{x}\in\mathbb{F}_{2}^{n}$, which is
sent through the memoryless channel $P_{Y|S,X}(y|s,x)$. The decoder
receives $\mathbf{y}=[y_{1},\ldots,y_{n}]$ where $y_{i}\in\mathcal{Y}$
(not necessarily binary), and outputs $\hat{\mathbf{m}}$. 
The input may also be subject to a cost constraint $\mathbf{E}[\sum_{i=1}^n c(s_i,x_i)] \le nD$, where $c : \mathcal{S}\times \mathbb{F}_2 \to [0,\infty)$.
The goal
is to design a coding scheme satisfying the cost constraint such that the error probability $\mathbf{P}(\mathbf{m}\neq\hat{\mathbf{m}})\to0$
as $n\to\infty$ when the message length is $k=\lfloor nR\rfloor$,
where $R>0$ is the rate.

We first briefly review the coding scheme based on Poisson functional representation given in~\cite{li2021unified} (also see~\cite{sfrl_trans}). Fix $P_{X|S}$. Let $\{Z_{\mathbf{m},\mathbf{x}} \}$ be the shared randomness between the encoder and the decoder, where $Z_{\mathbf{m},\mathbf{x}} \stackrel{iid}{\sim} \mathrm{Exp}(1)$ is the random bias for $\mathbf{m} \in\mathbb{F}_{2}^{k}$, $\mathbf{x}\in\mathbb{F}_{2}^{n}$. Given $\mathbf{m}$, $\mathbf{s}$, the encoder transmits $\mathbf{x} = \mathrm{argmax}_{\mathbf{x}} Z_{\mathbf{m},\mathbf{x}}^{-1} \prod_{i=1}^n P_{X|S} (x_i|s_i)$. The decoder finds $\hat{\mathbf{m}},\hat{\mathbf{x}}$ that maximize $Z_{\hat{\mathbf{m}},\hat{\mathbf{x}}}^{-1} \prod_{i=1}^n P_{X|Y} (\hat{x}_i|y_i)$, and outputs $\hat{\mathbf{m}}$. It was shown in~\cite{li2021unified} that this coding scheme achieves the rate $I(X;Y)-I(X;S)$, and hence can achieve the capacity given by the Gelfand-Pinsker theorem \cite{gelfand1980coding}.\footnote{While the Gelfand-Pinsker theorem involves
an auxiliary random variable $U$, we can 
treat $U$ as the channel input $X$ and apply the scheme on $U$.} 
It also attains the best known second-order bound in~\cite{scarlett2015dispersions}, and outperforms finite-blocklength results based on sub-codebooks given in~\cite{verdu2012nonasymp,yassaee2013oneshot,watanabe2015nonasymp}. The downside is that the shared randomness $\{Z_{\mathbf{m},\mathbf{x}} \}$ has a size exponential in $n$ and is unstructured, preventing efficient encoding and decoding. Our goal is to design a code with a linear structure that retains the advantage of~\cite{li2021unified}.

Consider the following coding scheme, which we call the \emph{weighted
parity-check codes with state}. 
Recall that the full parity-check matrix $\mathbf{H}$ is a uniformly
chosen random full-rank matrix.
The encoder observes $\mathbf{m}$
and $\mathbf{s}$ and uses the encoder codeword and parity bias
functions $\mathbf{p}_{e}(\mathbf{m},\mathbf{s})$, $\mathbf{q}_{e}(\mathbf{m},\mathbf{s})$
to obtain the codeword $\mathbf{x}$. The decoder uses the decoder
codeword and parity bias functions $\mathbf{p}_{d}(\mathbf{y})$,
$\mathbf{q}_{d}(\mathbf{y})$ to obtain $\hat{\mathbf{x}}$, and outputs
$\hat{\mathbf{m}}=[(\hat{\mathbf{x}}\mathbf{H}^{T})_{1},\ldots,\,(\hat{\mathbf{x}}\mathbf{H}^{T})_{k}]$.
We take
\begin{align}
\mathbf{p}_{e}(\mathbf{m},\mathbf{s}) & =[p_{e}(s_{1}),\ldots,p_{e}(s_{n})], &\mathbf{q}_{e}(\mathbf{m},\mathbf{s}) & =[\mathbf{m},\,\mathbf{q}],\nonumber\\
\mathbf{p}_{d}(\mathbf{y}) & =[p_{d}(y_{1}),\ldots,p_{d}(y_{n})], &\mathbf{q}_{d}(\mathbf{y}) & =[\frac{1}{2}\mathbf{1}^{k},\,\mathbf{q}], \label{eq:pq_state}
\end{align}
where $p_{e}:\mathcal{S}\to[0,1]$, $p_{d}:\mathcal{Y}\to[0,1]$ are
parameters of the encoder and decoder, and $\mathbf{q}=[q_{1},\ldots,q_{n-k}]$,
where $q_{i}\sim P_{Q}$ i.i.d., 
and $P_{Q}$ is a distribution over $[0,1]$ symmetric about $1/2$ (i.e., if $Q\sim P_{Q}$, then $1-Q \sim P_{Q}$),\footnote{The symmetry requirement  
is to avoid having a bias towards choosing $\mathbf{x}=0$, which gives $0$ for all parity-check bits. This requirement does not matter if the channel is symmetric.} called the \emph{parity
bias distribution}, which is a parameter of the code. While $\mathbf{q}$
and the full parity-check matrix $\mathbf{H}$ are regarded as common
randomness between the encoder and the decoder, they can agree on one
fixed choice of $\mathbf{q}$ and $\mathbf{H}$ via the standard derandomization
argument.

Expanding \eqref{eq:fH}, the encoding function is $\mathbf{x} = \mathrm{argmax}_{\mathbf{x}}  w_{[\mathbf{m},\mathbf{q}]}(\mathbf{x}\mathbf{H}^T) \prod_{i=1}^n \tilde{P}_{X|S}(x_i | s_i)$, where $\tilde{P}_{X|S}(x | s) = (p_e(s))^x (1-p_e(s))^{1-x}$.
Comparing this scheme with the aforementioned Poisson functional representation scheme, the random bias $Z_{\mathbf{m},\mathbf{x}}^{-1}$ is replaced with $w_{[\mathbf{m},\mathbf{q}]}(\mathbf{x}\mathbf{H}^T)$.
Considering that the length of the random vector $\mathbf{q}$ is only $n-k$ (compared to $2^{n+k}$ of $\{Z_{\mathbf{m},\mathbf{x}} \}$), it is impossible for $\{w_{[\mathbf{m},\mathbf{q}]}(\mathbf{x}\mathbf{H}^T)\}$ to have the same joint distribution as $\{Z_{\mathbf{m},\mathbf{x}} \}$. Nevertheless, we will show that under a certain condition on the distribution $P_Q$, this code can also achieve the capacity.

The nested linear code \cite{Zamir2002Nested,barron2003duality} can be regarded as
a special case of the weighted parity-check code with state, where there are $n-k-\tilde{k}$ parity-check bits that are fixed to zero (i.e., $q_i=0$), and $\tilde{k}$ unused parity-check bits ($q_i=1/2$), where $\tilde{k} \in \{0,\ldots,n-k\}$ is the dimension of each coset. 
This can be approximated by taking $P_{Q}(0)=P_{Q}(1)=(1-\gamma)/2$, $P_{Q}(1/2)=\gamma$, where $\gamma = \tilde{k}/(n-k)$, giving around $(n-k)P_Q(1/2)=\tilde{k}$ unused parity-check bits.\footnote{Here we have the same expected number of parity-check bits set to $1$ ($q_i=1$) as those set to $0$, due to the symmetry requirement on $P_Q$. Whether the parity-check bits are set to $0$ or $1$ does not matter in binary-Hamming information embedding~\cite{Zamir2002Nested,barron2003duality} due to symmetry.}

We now present the main result which shows that the WPC code is capacity achieving.
\begin{thm}
\label{thm:capacity}Assume $|\mathcal{S}|,|\mathcal{Y}|<\infty$. Fix any $P_{X|S}$, and let
$S\sim P_{S}$, $X|S\sim P_{X|S}$,
$Y|(S,X)\sim P_{Y|S,X}$. Consider the weighted parity-check code with
state, where $p_{e}(s)=P_{X|S}(1|s)$, $p_{d}(y)=P_{X|Y}(1|y)$, and $P_{Q}$ is a discrete distribution over $[0,1]$ symmetric about $1/2$ (i.e., $Q$ has the same distribution as $1-Q$) with finite support satisfying
\ifshortver
\else
\begin{equation}
\mathbf{E}[H_{b}(Q)]=\frac{1-H(X|S)}{1-R}.\label{eq:ehbq}
\end{equation}
\fi
For any $R<I(X;Y)-I(X;S)$, as $n\to\infty$, the probability
of error of the code tends to $0$, and the empirical joint distribution of $\{(s_{i},x_{i})\}_{i=1,\ldots,n}$
tends to $P_{S}P_{X|S}$ in probability.
\end{thm}

To prove the theorem, we require the following technical lemma that gives a sufficient condition for the probability of error to tend to $0$. 
\ifshortver
\else
The proof is given later in this section. 
\fi

\begin{lem}
\label{lem:state}Consider the weighted parity-check code with state,
where $|\mathcal{S}|,|\mathcal{Y}|<\infty$, and $P_{Q}$ is a discrete distribution over $[0,1]$ symmetric about $1/2$ with finite support. Let
$S\sim P_{S}$, $X|S\sim P_{X|S}$,
$Y|(S,X)\sim P_{Y|S,X}$, and let $Q,V$ be auxiliary random variables following $Q\sim P_{Q}$, $V\in \{0,1\}$, $V|Q\sim P_{V|Q}$, where $(P_{X|S},P_{V|Q})$ is the 
minimizer of 
\begin{align}
 & \mathbf{E}\left[H_{b}(X,p_{e}(S))\right]+(1-R)\mathbf{E}\left[H_{b}(V,Q)\right],\label{eq:ehb_min}
\end{align}
where $H_{b}$ is the binary cross entropy function \eqref{eq:hb},
subject to
\begin{equation}
H(X|S)+(1-R)H(V|Q)\ge1.\label{eq:ehb_min_cons}
\end{equation}
If the minimizer of~\eqref{eq:ehb_min} is unique, and for all $P_{\tilde{X}|Y}$, $P_{\tilde{V}|Q}$ 
satisfying
\begin{equation}
H(\tilde{X}|Y)+(1-R)H(\tilde{V}|Q)\ge1-R,\label{eq:hxy_hvq}
\end{equation}
we have
\begin{align}
 & \mathbf{E}[H_{b}(\tilde{X},p_{d}(Y))]+(1-R)\mathbf{E}[H_{b}(\tilde{V},Q)]\nonumber\\
 & >\mathbf{E}[H_{b}(X,p_{d}(Y))]+(1-R)\mathbf{E}[H_{b}(V,Q)],\label{eq:dec_ineq}
\end{align}
then as $n\to\infty$, the probability of error of the code tends
to $0$, and the empirical joint distribution of $\{(s_{i},x_{i})\}_{i=1,\ldots,n}$
tends to $P_{S}P_{X|S}$ in probability.
\end{lem}

To prove Theorem~\ref{thm:capacity}, 
for any fixed
$\lambda\ge0$, the maximizer of $H(X|S)-\lambda\mathbf{E}[H_{b}(X,p_{e}(S))]$
is the tilted distribution
\[
P_{X|S}(1|s)=(p_{e}(s))^{\lambda} / ((p_{e}(s))^{\lambda}+(1-p_{e}(s))^{\lambda}).
\]
Similarly, the maximizer of $H(V|Q)-\lambda\mathbf{E}[H_{b}(V,Q)]$
is
$P_{V|Q}(1|q)= q^{\lambda}/( q^{\lambda}+(1-q)^{\lambda})$.
Hence, to solve the minimization problem \eqref{eq:ehb_min}, we can choose $\lambda$ such that the equality in \eqref{eq:ehb_min_cons}
holds. In particular, if equality holds when $\lambda=1$, then the WPC
code can achieve the capacity of the channel.
We now present the proof of Theorem~\ref{thm:capacity}.

\begin{IEEEproof}[Proof of Theorem~\ref{thm:capacity}]
Note that $P_{X|S}(1|s)=p_{e}(s)$, $P_{V|Q}(1|q)=q$ is the unique maximizer of $H(X|S)-\mathbf{E}[H_{b}(X,p_{e}(S))]+(1-R)(H(V|Q)-\mathbf{E}[H_{b}(V,Q)])$. By~\eqref{eq:ehbq}, we can deduce that $(P_{X|S},P_{V|Q})$ is the unique minimizer of \eqref{eq:ehb_min}.
It remains to check~\eqref{eq:dec_ineq}.
We have $\mathbf{E}[H_{b}(Q)]=H(V|Q)$, $\mathbf{E}[H_{b}(X,p_{d}(Y))]=H(X|Y)$, and $\mathbf{E}[H_{b}(\tilde{X},p_{d}(Y))]\ge H(\tilde{X}|Y)$.
If $R<H(X|S)-H(X|Y)$, then
\ifshortver
\else
\begin{align*}
 & \mathbf{E}[H_{b}(\tilde{X},p_{d}(Y))]+(1-R)\mathbf{E}[H_{b}(\tilde{V},Q)]\\
 & \ge H(\tilde{X}|Y)+(1-R)\mathbf{E}[H_{b}(\tilde{V},Q)]\\
 & \stackrel{(a)}{\ge} 1-R+(1-R)\big(\mathbf{E}[H_{b}(\tilde{V},Q)] - H(\tilde{V}|Q)\big)\\
 & >1\! -\! H(X|S)\! +\! H(X|Y)+(1 \! -\! R)\big(\mathbf{E}[H_{b}(\tilde{V}\!,Q)]\! -\! H(\tilde{V}|Q)\!\big)\\
 & \stackrel{(b)}{=}H(X|Y) +(1-R)\big(H(V|Q) + \mathbf{E}[H_{b}(\tilde{V},Q)] - H(\tilde{V}|Q)\big)\\
 & =\mathbf{E}[H_{b}(X,p_{d}(Y))]\\
 &\;\;\;\;+(1-R)\big(H(V|Q)+\mathbf{E}[H_{b}(\tilde{V},Q)]-H(\tilde{V}|Q)\big)\\
 & \stackrel{(c)}{\ge}\mathbf{E}[H_{b}(X,p_{d}(Y))]+(1-R)\mathbf{E}[H_{b}(V,Q)],
\end{align*}
where (a) is by~\eqref{eq:hxy_hvq}, (b) is by~\eqref{eq:ehbq}, and (c) is because $P_{V|Q}(1|q)=q$
maximizes $H(V|Q)-\mathbf{E}[H_{b}(V,Q)]$. The result follows from
Lemma~\ref{lem:state}.
\fi
\end{IEEEproof}

\medskip{}

Note that Theorem~\ref{thm:capacity} can be applied on asymmetric channels. In particular, for asymmetric channels without state (i.e., $S=\emptyset$), we take $p_e(s)=P_X(1)$ where $P_X$ is the capacity-achieving input distribution, $p_d(y)=P_{X|Y}(1|Y)$, and $P_Q$ with $\mathbf{E}[H_{b}(Q)]=(1-H(X))/(1-R)$. Nevertheless, Theorem~\ref{thm:capacity} only applies when $X$ (or the auxiliary in the Gelfand-Pinsker theorem) is binary. For $X$ in a general finite field $\mathbb{F}_l$, we apply the extension discussed in Remark~\ref{rem:finitefield}. Now $P_Q$ is a distribution over $\Delta_{\mathbb{F}_l}$, the probability simplex over $\mathbb{F}_l$, and we require $\mathbf{E}[H_{\mathbb{F}_l}(Q)]=(\log l -H(X))/(1-R)$ instead of \eqref{eq:ehbq}, where $H_{\mathbb{F}_l}(Q)$ denotes the entropy of the probability vector $Q$ (not the entropy of the random variable $Q$). 
\ifshortver
\else
The proofs of the extensions of Theorem~\ref{thm:capacity} to general $X$ are similar to the binary case and are omitted.
\fi

Using Theorem~\ref{thm:capacity}, to achieve the capacity with
the input distribution $P_{X|S}$, we construct the
parity bias distribution $P_{Q}$ symmetric about $1/2$ so that \eqref{eq:ehbq} holds.
We discuss the following choices of $P_{Q}$.
\begin{itemize}
\item (Threshold) Take $P_{Q}(0)=P_{Q}(1)=(1-\gamma)/2$, $P_{Q}(1/2)=\gamma$, where
\begin{align}
    \gamma=\frac{1-H(X|S)}{1-R}.\label{eq:thres_gamma}
\end{align}
This is essentially equivalent
to the nested linear code.
\item (Constant) Take $P_{Q}(c)=P_{Q}(1-c)=1/2$, where $c=H_{b}^{-1}((1-H(X|S))/(1-R))$.
\item (Linear) Take $P_{Q}$ to be the uniform distribution $\mathrm{Unif}[0,1]$.\footnote{Lemma \ref{lem:state} requires $P_{Q}$ to be discrete, though
this is mostly a technical requirement of the proof.} This may not achieve the capacity, but has the advantage of being
``universal'' in the sense that the decoder does not need to know
$P_{S}$ or $p_{e}$.
\item (Threshold linear) 
Construct $P_{Q}$ using the cumulative distribution function
\begin{equation}
F_Q(t) := \begin{cases}
0 & \mathrm{if}\; t < 0 \\
\max\{\theta / 2 ,\,0\} & \mathrm{if}\; 0 \le t < |\theta| / 2 \\
t & \mathrm{if}\; |\theta| / 2 \le t < 1 - |\theta| / 2 \\
1-\max\{\theta / 2 ,\,0\} & \mathrm{if}\; 1- |\theta| / 2 \le t < 1\\
1 & \mathrm{if}\; t \ge 1, \\
\end{cases}\label{eq:thres_linear}
\end{equation}
where $\theta \in [-1,1]$ is chosen such that \eqref{eq:ehbq} holds.
This combines the linear method for $t$ close to $1/2$, and the threshold method for smaller and larger $t$'s.
This method achieves a better performance than the nested linear code, which will be discussed in Section~\ref{subsec:info_embed}.
\end{itemize}
In practice, to reduce the randomness in the code construction, 
entries of $\mathbf{q}=[q_{1},\ldots,q_{n-k}]$ is not taken i.i.d on $P_{Q}$, instead, 
\[
q_{i}=F_{Q}^{-1}\left(\frac{i-3/4}{n-k}\right),
\]
where $F_{Q}^{-1}$ is the inverse cumulative distribution function
of $P_{Q}$. This ensures that the empirical distribution of $\{q_{i}\}$
is close to $P_{Q}$. 
Note that we use $i-3/4$ instead of $i-1/2$ since $|(i-3/4)/(n-k) - 1/2|$ is closer to being uniformly distributed over $[0,1/2]$.
This also allows us to recover the nested linear code as a special case exactly.

\ifshortver
\else
We now prove Lemma \ref{lem:state}. We require the following lemma.
\begin{lem}
\label{lem:ab_bd}Consider the finite field $\mathbb{F}_{l}$. Let
$A,B\subseteq\mathbb{F}_{l}^{n}\backslash\{\mathbf{0}\}$ be arbitrary
subsets. Let $\mathbf{H}\in\mathbb{F}_{l}^{n\times n}$ be a uniformly
randomly chosen full-rank matrix. Let 
\[
\theta:=\log_{l}\frac{|A||B|}{n}-n-1.
\]
If $\theta^{-1}\ln\theta\ge l^{-n}|A|$, we have 
\[
\mathbf{P}\left(A\cap\mathbf{H}B=\emptyset\right)\le\frac{1+\ln\theta}{\theta},
\]
where $\mathbf{H}B:=\{\mathbf{H}\mathbf{x}:\,\mathbf{x}\in B\}$. 
\end{lem}
The proof is in Appendix \ref{lem:ab_bd}. A direct corollary is that
if $\log_{l}|A|=nR_{1}+o(n)$, $\log_{l}|B|=nR_{2}+o(n)$, $R_{1}+R_{2}>1$,
then $\mathbf{P}(A\cap\mathbf{H}B=\emptyset)\to0$ as $n\to\infty$.

We are ready to prove Lemma \ref{lem:state}.
\begin{IEEEproof}
[Proof of Lemma \ref{lem:state}] We assume that the encoder and
the decoder, instead of using \eqref{eq:fH}, use the following function
\begin{equation}
f_{\mathbf{H},\mathbf{b}}(\mathbf{p},\mathbf{q}):=\mathrm{argmax}_{\mathbf{x}\in\mathbb{F}_{2}^{n}}\,w_{\mathbf{p}}(\mathbf{x})w_{\mathbf{q}}(\mathbf{x}\mathbf{H}^{T}\oplus \mathbf{b}),\label{eq:fxHb}
\end{equation}
where $\mathbf{b}\sim\mathrm{Unif}(\mathbb{F}_{2}^{n})$ is shared
by the encoder and the decoder. This flips the parity-check bits randomly, which does not affect the performance
of the code due to the symmetry of $P_{Q}$, though it eliminates the need of treating the zero vector separately
and simplifies the proof. Fix $0<\epsilon_{1},\ldots,\epsilon_{6}<1$.
Fix any typical sequences $\mathbf{s}\in\mathcal{T}_{\epsilon_{1}}^{(n)}(P_{S})$, $\mathbf{q}\in\mathcal{T}_{\epsilon_{1}}^{(n)}(P_{Q})$. Asymptotic
equipartition property gives
\[
|\mathcal{T}_{2\epsilon_{1}}^{(n)}(\bar{P}_{X|S}^{(\epsilon_{2})}|\mathbf{s})|\ge(1-2\epsilon_{1})2^{n(\epsilon_{2}+(1-\epsilon_{2})H(X|S)-\delta(\epsilon_{1}))},
\]
\[
|\mathcal{T}_{2\epsilon_{1}}^{((1-R)n)}(\bar{P}_{V|Q}^{(\epsilon_{2})}|\mathbf{q})|\ge(1-2\epsilon_{1})2^{(1-R)n(\epsilon_{2}+(1-\epsilon_{2})H(V|Q)-\delta(\epsilon_{1}))},
\]
where $\delta(\epsilon_{1})\to0$ as $\epsilon_{1}\to0$,  $\bar{P}_{X|S}^{(\epsilon_{2})}(x|s):=\epsilon_{2}/2+(1-\epsilon_{2})P_{X|S}(x|s)$ is a mixture between $P_{X|S}$ and $\mathrm{Bern}(1/2)$, and $\bar{P}_{V|Q}^{(\epsilon_{2})}$ is defined similarly.
Note that 
\begin{equation*}
H_{(S,X)\sim P_S\bar{P}_{X|S}^{(\epsilon_{2})}}(X|S) \ge \epsilon_{2}+(1-\epsilon_{2})H(X|S)
\end{equation*}
due to concavity of entropy.
Since $H(X|S)+(1-R)H(V|Q)\ge1$ by \eqref{eq:ehb_min_cons}, we have
\begin{align*}
1< & \epsilon_{2}+(1-\epsilon_{2})H(X|S)-\delta(\epsilon_{1})\\
 & \;\;+(1-R)(\epsilon_{2}+(1-\epsilon_{2})H(V|Q)-\delta(\epsilon_{1}))
\end{align*}
for any fixed $0<\epsilon_{2}<1$ and for $\epsilon_{1}$ small enough.
By Lemma \ref{lem:ab_bd}, there exist $\mathbf{x}^{*}\in\mathcal{T}_{2\epsilon_{1}}^{(n)}(\bar{P}_{X|S}^{(\epsilon_{2})}|\mathbf{s})$,
$\mathbf{v}^{*}\in\mathcal{T}_{2\epsilon_{1}}^{((1-R)n)}(\bar{P}_{V|Q}^{(\epsilon_{2})}|\mathbf{q})$
satisfying $\mathbf{x}^{*}\mathbf{H}^{T}\oplus \mathbf{b}=[\mathbf{m},\mathbf{v}^{*}]$
with probability approaching $1$ as $n\to\infty$. Letting $\epsilon_{1},\epsilon_{2}\to0$,
there exist $\mathbf{x}^{*}\in\mathcal{T}_{\epsilon_{3}}^{(n)}(P_{X|S}|\mathbf{s})$,
$\mathbf{v}^{*}\in\mathcal{T}_{\epsilon_{3}}^{((1-R)n)}(P_{V|Q}|\mathbf{q})$ for some arbitrarily small $\epsilon_3$.

We then show that the sequence chosen by the encoder (maximizer of
\eqref{eq:fxHb}) can only be one of such typical sequences. Fix any
$\mathbf{m}$. Let $\mathbf{x}\sim\mathrm{Unif}(\mathbb{F}_{2}^{n})$, and $\mathbf{m}',\mathbf{v}$ be such that $[\mathbf{m}',\mathbf{v}]=\mathbf{x}\mathbf{H}^{T}\oplus \mathbf{b}$
(note that $[\mathbf{m}',\mathbf{v}]\sim\mathrm{Unif}(\mathbb{F}_{2}^{n})$
is independent of $\mathbf{x}$ since $\mathbf{H},\mathbf{b}$ are also
random). By Sanov's theorem, 
\begin{align}
 & \mathbf{P}\big(\mathbf{m}'=\mathbf{m},\,H(\hat{X}|\hat{S})+(1-R)H(\hat{V}|\hat{Q})\le1-\epsilon_{4}\big)\label{eq:PxH}\\
 & =\mathbf{P}\big(\mathbf{m}'=\mathbf{m},\,D(P_{\hat{X}|\hat{S}}\Vert\mathrm{Bern}(1/2)|\hat{S})\nonumber \\
 & \;\;\;\;+(1-R)D(P_{\hat{V}|\hat{Q}}\Vert\mathrm{Bern}(1/2)|\hat{Q})\ge1-R+\epsilon_{4}\big)\nonumber \\
 & =\mathbf{P}\big(\mathbf{m}'=\mathbf{m}\big) \mathbf{P}\big(D(P_{\hat{X}|\hat{S}}\Vert\mathrm{Bern}(1/2)|\hat{S})\nonumber \\
 & \;\;\;\;+(1-R)D(P_{\hat{V}|\hat{Q}}\Vert\mathrm{Bern}(1/2)|\hat{Q})\ge1-R+\epsilon_{4}\big)\nonumber \\
 & \le2^{-nR} \cdot 2^{-n(1-R+\epsilon_{4}-\delta(\epsilon_{1}))+o(n)}\nonumber \\
 & =2^{-n(1+\epsilon_{4}-\delta(\epsilon_{1}))+o(n)},\nonumber 
\end{align}
where $(\hat{X},\hat{S})\sim\hat{P}_{\mathbf{x},\mathbf{s}}$, $(\hat{V},\hat{Q})\sim\hat{P}_{\mathbf{v},\mathbf{q}}$.
Hence, for small enough $\epsilon_1,\epsilon_4$, with probability approaching
$1$, there does not exist $\mathbf{x}\in\mathbb{F}_{2}^{n}$
satisfying the condition in \eqref{eq:PxH}. Let $\mathbf{x}$ satisfy $\mathbf{x}\mathbf{H}^{T}\oplus \mathbf{b}=[\mathbf{m},\mathbf{v}]$,
$H(\hat{X}|\hat{S})+(1-R)H(\hat{V}|\hat{Q})>1-\epsilon_{4}$ and within
$\epsilon_{4}$ from optimal in \eqref{eq:ehb_min}, i.e.,
\begin{align}
 & \mathbf{E}\big[H_{b}(\hat{X},p_{e}(\hat{S}))\big]+(1-R)\mathbf{E}\big[H_{b}(\hat{V},\hat{Q})\big]\label{eq:x_alt_lhs}\\
 & \le\mathbf{E}\big[H_{b}(X,p_{e}(S))\big]+(1-R)\mathbf{E}\big[H_{b}(V,Q)\big]+\epsilon_{4}.\nonumber 
\end{align}
Note that the left hand side \eqref{eq:x_alt_lhs} above is the negative
logarithm of \eqref{eq:fH}, and the right hand side (without the
``$\epsilon_{4}$'') is arbitrarily close to \eqref{eq:x_alt_lhs}
evaluated on $\mathbf{x}^{*},\mathbf{q}^{*}$, so the above must be
satisfied if $\mathbf{x}$ is chosen by the encoder instead of $\mathbf{x}^{*}$.
By the uniqueness of minimizer in \eqref{eq:ehb_min} and continuity,
we have $\mathbf{x}\in\mathcal{T}_{\epsilon_{5}}^{(n)}(P_{X|S}|\mathbf{s})$
and $\mathbf{v}\in\mathcal{T}_{\epsilon_{5}}^{((1-R)n)}(P_{V|Q}|\mathbf{q})$,
for fixed $\epsilon_{5}$ as $\epsilon_{4}\to0$. Hence, the sequence
$\mathbf{x}$ chosen by the encoder must be conditionally typical
given $\mathbf{s}$. We can assume $(\mathbf{s},\mathbf{x},\mathbf{y})\in\mathcal{T}_{\epsilon_{5}}^{(n)}(P_{S,X,Y})$,
$(\mathbf{v},\mathbf{q})\in\mathcal{T}_{\epsilon_{5}}^{((1-R)n)}(P_{V,Q})$
are jointly typical for some arbitrarily small $\epsilon_{5}$.

It is left to bound the error probability. 
Note that the encoder does not depend on the whole $(\mathbf{H},\mathbf{b})$. Instead, it only depends on
\begin{equation}
\mathcal{C}_{\mathbf{H},\mathbf{b}}(\mathbf{m}) := \{(\mathbf{x},\mathbf{v}):\, [\mathbf{m},\mathbf{v}]=\mathbf{x}\mathbf{H}^{T}\oplus \mathbf{b}\},\label{eq:enc_dep}
\end{equation}
i.e., the set of sequences that correspond to the message $\mathbf{m}$.
Let $\mathcal{X}_{\mathbf{H},\mathbf{b}}(\mathbf{m}) := \{\mathbf{x}:\,\exists \mathbf{v}.\,(\mathbf{x},\mathbf{v}) \in \mathcal{C}_{\mathbf{H},\mathbf{b}}(\mathbf{m})\}$.
Let $\tilde{\mathbf{x}}\sim\mathrm{Unif}(\mathbb{F}_{2}^{n})$, and $\tilde{\mathbf{m}},\tilde{\mathbf{v}}$ be such that $\tilde{\mathbf{x}}\mathbf{H}^{T} \oplus \mathbf{b}=[\tilde{\mathbf{m}},\tilde{\mathbf{v}}]$. If $\tilde{\mathbf{x}} \in \mathcal{X}_{\mathbf{H},\mathbf{b}}(\mathbf{m})$, then the value of $\tilde{\mathbf{v}}$ is determined by $\tilde{\mathbf{x}}, \mathcal{C}_{\mathbf{H},\mathbf{b}}(\mathbf{m})$. Otherwise, $\tilde{\mathbf{v}}$ is uniform over $\mathbb{F}_{2}^{n-k}$ (since $\mathcal{C}_{\mathbf{H},\mathbf{b}}(\mathbf{m})$ only fixes the values of the affine mapping $\mathbf{x} \mapsto \mathbf{x}\mathbf{H}^{T}\oplus \mathbf{b}$ over an affine subspace, there are enough degrees of freedom in $\mathbf{H},\mathbf{b}$ to bring any $\tilde{\mathbf{x}}$ outside of that affine subspace to any $\tilde{\mathbf{v}}$). Therefore, we have the following conditional distribution
\begin{align}
&(\tilde{\mathbf{x}},\tilde{\mathbf{v}}) \, \big|\, \mathcal{C}_{\mathbf{H},\mathbf{b}}(\mathbf{m}) \,\sim\, \nonumber\\ 
&\;\;\; 2^{-k}\mathrm{Unif}(\mathcal{C}_{\mathbf{H},\mathbf{b}}(\mathbf{m})) + (1-2^{-k})\mathrm{Unif}(\mathcal{X}^c_{\mathbf{H},\mathbf{b}}(\mathbf{m}) \times \mathbb{F}_{2}^{n-k}), \label{eq:xv_dist}
\end{align}
where $\mathcal{X}^c$ is the complement of $\mathcal{X}$.
Let $(\hat{\tilde{X}},\hat{S})\sim\hat{P}_{\tilde{\mathbf{x}},\mathbf{s}}$,
$(\hat{\tilde{V}},\hat{Q})\sim\hat{P}_{\tilde{\mathbf{v}},\mathbf{q}}$.
We have
\begin{align}
 & \mathbf{P}\Big(\tilde{\mathbf{x}} \notin \mathcal{X}_{\mathbf{H},\mathbf{b}}(\mathbf{m}),\, H(\hat{\tilde{X}}|\hat{Y})+(1-R)H(\hat{\tilde{V}}|\hat{Q}) \nonumber \\
 &\;\;\;\;\;\;\;\le1-R-\epsilon_{6} \,\Big|\, \mathcal{C}_{\mathbf{H},\mathbf{b}}(\mathbf{m}),\mathbf{y},\mathbf{q}\Big)\label{eq:Pxth}\\
 & \stackrel{(a)}{=} \mathbf{P}_{(\tilde{\mathbf{x}},\tilde{\mathbf{v}}) \sim \mathrm{Unif}(\mathbb{F}_{2}^{n} \times \mathbb{F}_{2}^{n-k})}\Big(\tilde{\mathbf{x}} \notin \mathcal{X}_{\mathbf{H},\mathbf{b}}(\mathbf{m}),\,  \nonumber \\
 &\;\;\;\;\;\;\;\,H(\hat{\tilde{X}}|\hat{Y})+(1-R)H(\hat{\tilde{V}}|\hat{Q}) \le1-R-\epsilon_{6} \Big)\nonumber\\
 & \le \mathbf{P}_{(\tilde{\mathbf{x}},\tilde{\mathbf{v}}) \sim \mathrm{Unif}(\mathbb{F}_{2}^{n} \times \mathbb{F}_{2}^{n-k})}\big(D(P_{\hat{\tilde{X}}|\hat{Y}}\Vert\mathrm{Bern}(1/2)|\hat{Y})\nonumber \\
 & \;\;\;\;+(1-R)D(P_{\hat{\tilde{V}}|\hat{Q}}\Vert\mathrm{Bern}(1/2)|\hat{Q})\ge1+\epsilon_{6}\big)\nonumber \\
 & \stackrel{(b)}{\le} 2^{-n(1+\epsilon_{6}-\delta(\epsilon_{5}))+o(n)},\nonumber 
\end{align}
where (a) is due to~\eqref{eq:xv_dist} since the distribution in (a) and the distribution~\eqref{eq:xv_dist} coincide when $\tilde{\mathbf{x}} \notin \mathcal{X}_{\mathbf{H},\mathbf{b}}(\mathbf{m})$, and (b) is by Sanov's theorem.
Hence with probability approaching
$1$, there does not exist $\tilde{\mathbf{x}}\notin \mathcal{X}_{\mathbf{H},\mathbf{b}}(\mathbf{m})$
satisfying the condition in \eqref{eq:Pxth}. Let $\tilde{\mathbf{x}} \notin \mathcal{X}_{\mathbf{H},\mathbf{b}}(\mathbf{m})$ and $\tilde{\mathbf{v}}$ satisfy $H(\hat{\tilde{X}}|\hat{Y})+(1-R)H(\hat{\tilde{V}}|\hat{Q})>1-R-\epsilon_{6}$
and
\begin{align*}
 & \mathbf{E}\big[H_{b}(\hat{\tilde{X}},p_{d}(\hat{Y}))\big]+(1-R)\mathbf{E}\big[H_{b}(\hat{\tilde{V}},\hat{Q})\big]\\
 & \le\mathbf{E}\big[H_{b}(X,p_{d}(Y))\big]+(1-R)\mathbf{E}\big[H_{b}(V,Q)\big]+\epsilon_{6}.
\end{align*}
The above inequality must be satisfied if $\tilde{\mathbf{x}}$ is chosen by
the decoder instead of $\mathbf{x}$. Since the gap in the inequality
in~\eqref{eq:dec_ineq} is bounded away from $0$ (can be proved by continuity),
this contradicts~\eqref{eq:dec_ineq} when $\epsilon_{6}\to0$. Hence, with
high probability, the decoder cannot choose any sequence $\tilde{\mathbf{x}} \notin \mathcal{X}_{\mathbf{H},\mathbf{b}}(\mathbf{m})$. 
\end{IEEEproof}
\medskip{}
\fi

Theorem \ref{thm:capacity} shows that the WPC codes achieve the capacity of an arbitrary binary channel with state. The generality of WPC codes is one of its advantages over the nested linear codes (which only apply to symmetric channels). Another advantage -- a lower error rate -- cannot be shown by Theorem \ref{thm:capacity}. To compare the performance of the codes, we will perform experiments in Section~\ref{subsec:info_embed}.

\subsection{Example -- Binary-Hamming Information Embedding}\label{subsec:info_embed}

We consider information embedding in binary-Hamming case~\cite{barron2003duality}, where we consider a channel with state $s_i \stackrel{iid}{\sim} \mathrm{Bern}(1/2)$ and $P_{Y|S,X}(1|s,0)=P_{Y|S,X}(0|s,1) = \beta$ (i.e., $X \to Y$ is $\mathrm{BSC}(\beta)$). Moreover, we have an expected cost (or distortion) constraint $\mathbf{E}[|\{i\in \{1,\ldots,n\}:\, x_i \neq s_i\}|] \le nD$, where $0<D<1$ is the maximum average cost per symbol. 
The goal is to achieve the optimal tradeoff between the rate $R$, the error probability $\mathbf{P}(\mathbf{m} \neq \hat{\mathbf{m}})$, and the expected cost $\mathbf{E}[|\{i:\, x_i \neq s_i\}|]$. 
As shown in~\cite{barron2003duality}, the capacity $C$ of this setting is given by $C=\mathfrak{E}[g(D)]$ where \begin{equation*}
g(D)=\left\{
\begin{aligned}
&0, & & \mathrm{if}\; 0\leq D<\beta,  \\
&H(D)-H(\beta), &  & \mathrm{if}\; \beta\leq D\leq 1/2,
\end{aligned}
\right.
\end{equation*}
and $\mathfrak{E}[g(D)]$ denotes the upper concave envelope of $g(D)$. 

\begin{figure}
	\centering
	\ifshortver
\else
	\includegraphics[width=0.9\linewidth]{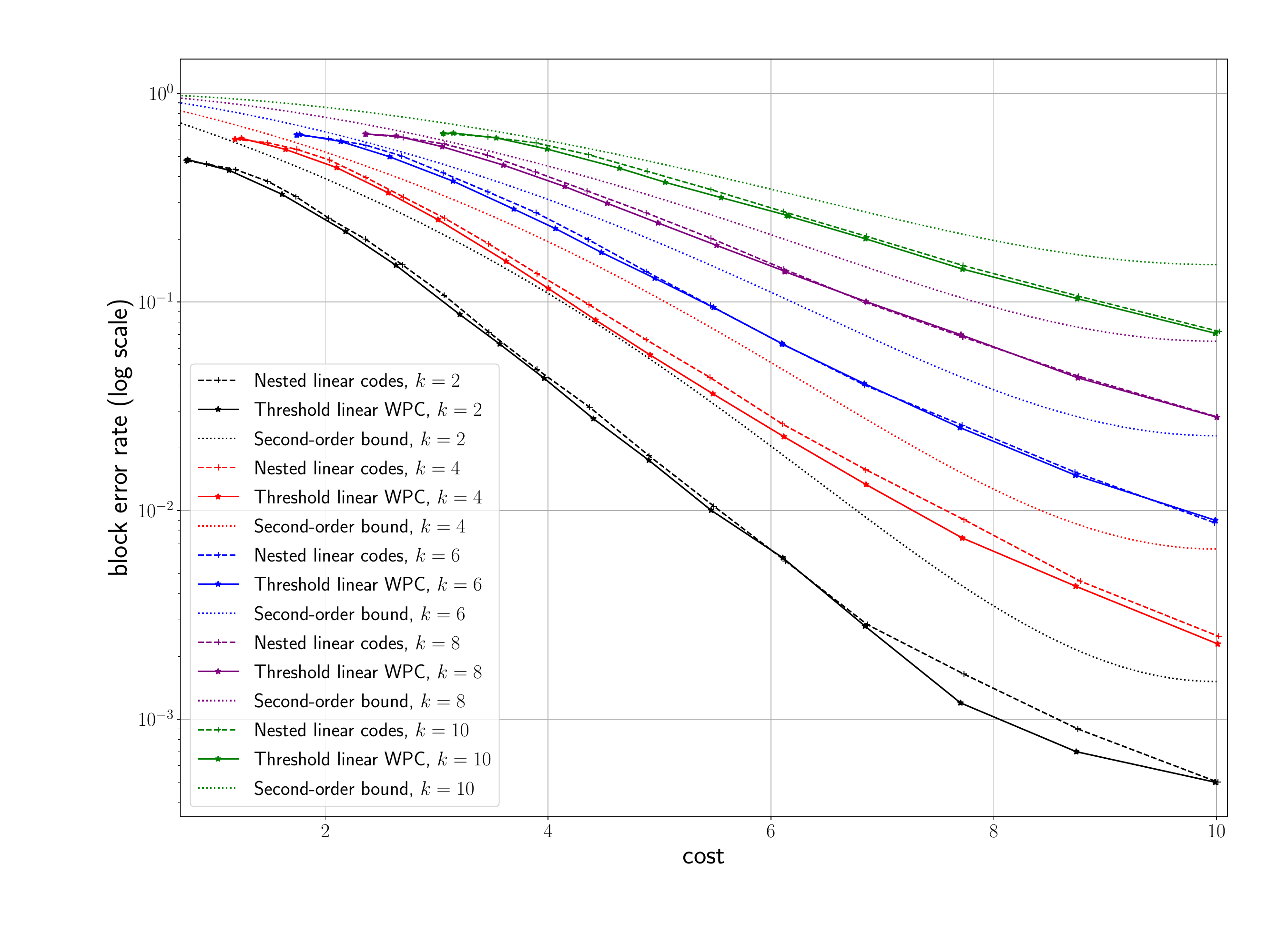}
	\fi
	\caption{Performance evaluation with $n=20$, $\beta=0.05$. Each data point is computed by $2\times 10^4$ trials.}
	\label{fig::info_embed}
	\vspace{-8pt}
\end{figure}

\begin{figure}
	\centering
	\includegraphics[width=0.9\linewidth]{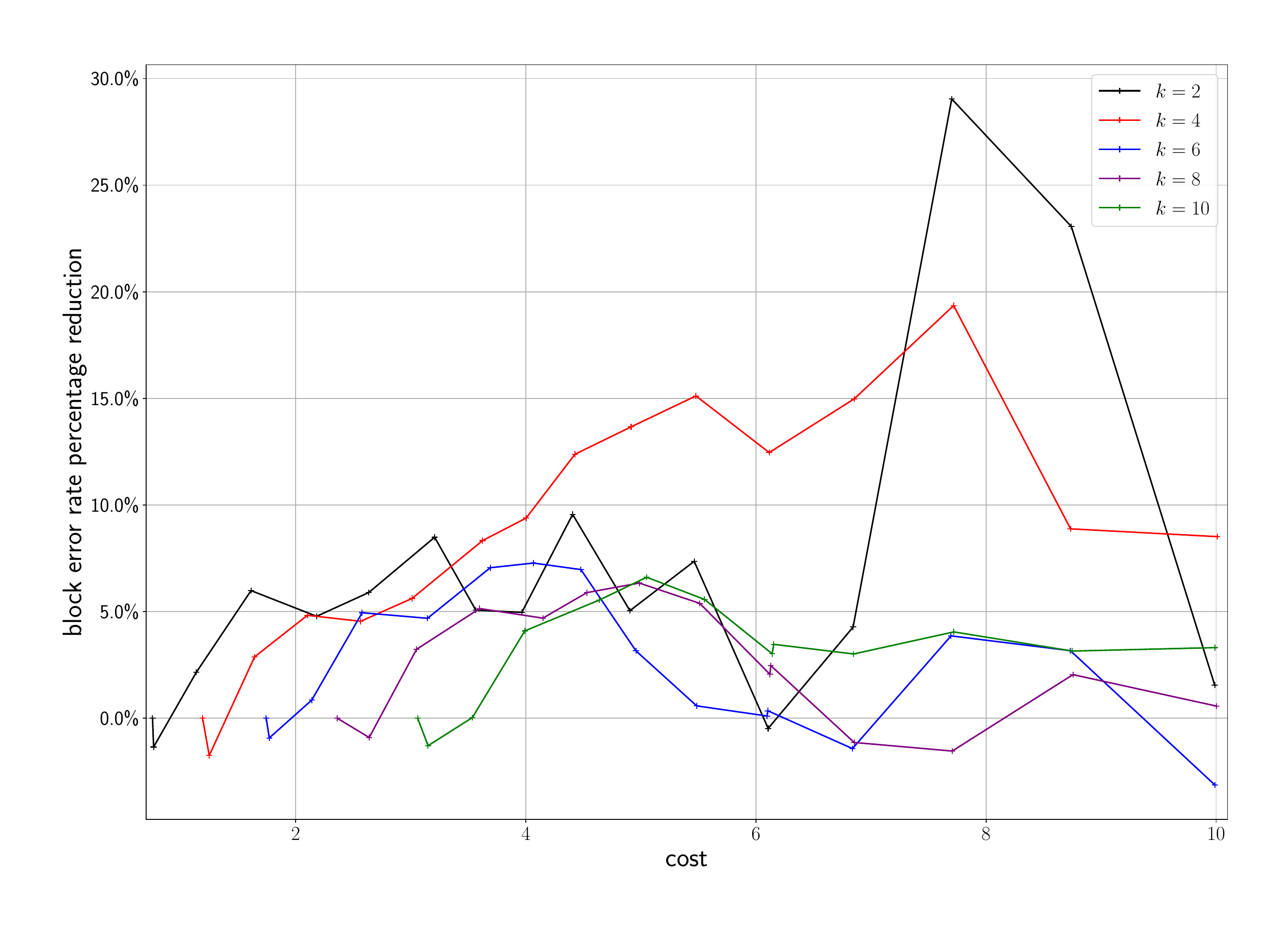}
	\caption{Performance evaluation on the percentage reduction of the 
error rates for the weighted parity-check codes compared to nested linear codes, 
with $n=20$, $\beta=0.05$. Linear interpolation is used on the data points of the nested linear codes for calculation. 
Each data point is computed by $2\times 10^4$ trials.}
	\label{fig::info_embed2}
	\vspace{-8pt}
\end{figure}

We compare the weighted parity-check code to the nested linear code\footnote{The parity-check matrices are generated uniformly at random with full rank. The encoder and decoder perform tie-breaking according to lexicographical order of the codeword.} \cite{Zamir2002Nested,barron2003duality} for the information embedding setting with $n=20$, $\beta=0.05$, and $k=2,4,6,8,10$. For the WPC code, we use $p_e(s)=\alpha^{1-s}(1-\alpha)^s$ so that $S\to X$ is approximately $\mathrm{BSC}(\alpha)$, and the expected cost $\mathbf{E}[|\{i:\, x_i \neq s_i\}|] \approx n \alpha$, where $0\le \alpha \le 1/2$ is called the \emph{cost parameter}. We use $p_d(y)=\beta^{1-y}(1-\beta)^y$ (the posterior distribution of $X$ given $Y$) and the threshold linear method \eqref{eq:thres_linear} for $P_Q$.
The plot of the block error rate against the average cost is shown in Figure~\ref{fig::info_embed}. 
The plot of the percentage reduction of the block error rate of the WPC code compared to the nested linear code (i.e., $1-(\text{error rate of WPC})/(\text{error rate of nested linear})$) is shown in Figure~\ref{fig::info_embed2}.\footnote{The percentage reduction is calculated by linear interpolating the data points of nested linear codes and comparing them with the data points of WPC. 
Other interpolation methods, e.g., cubic interpolation, give similar results.}
For the nested linear code, we perform simulation for each choice of the coset dimension parameter $\tilde{k}=0,\ldots,n-k$, where each $\tilde{k}$ corresponds to one point in the plot. For the WPC code, we perform simulation for each of the $26$ choices of the cost parameter $\alpha$ evenly spaced in $[0,1/2]$, i.e., the interval between choices of $\alpha$ is $0.02$, where each $\alpha$ corresponds to one point in the plot. In the simulation, we have performed $2\times 10^4$ trials for each data point in the plot to obtain the block error rate and the average cost.
We also plot the second-order achievability bound in~\cite[Theorem 1]{scarlett2015dispersions}.

From Figures~\ref{fig::info_embed} and~\ref{fig::info_embed2}, we can observe that the WPC code outperforms the nested linear code for most pairs of message lengths and costs.
WPC reduces the error rate by at most $27.3\%$ when $k=2$ and cost is about $7.7$.
While the reduction appears small, since the nested linear code is already capacity-achieving, it would be unrealistic to expect an improvement by an order of magnitude.

We only perform the experiment on the threshold linear option \eqref{eq:thres_linear} for the parity bias $P_Q$. The constant and linear option have significantly poorer performance. Intuitively, it is undesirable to have a small nonzero $q_i$. Compared to the encoder, the decoder has a stronger belief on the codeword $\mathbf{x}$, allowing it to ignore the small value of $q_i$ and decode to a codeword with the $i$-th parity-check bit $(\hat{\mathbf{x}}\mathbf{H}^{T})_{k+i}=1$. Therefore, having a small nonzero $q_i$ makes it unlikely for the encoder to select 1 as the $i$-th parity-check bit, but not as unlikely for the decoder, resulting in the existence of codewords that cannot be selected by the encoder, but can be selected by the decoder. Therefore, a ``threshold'' method where each $q_i$ is either zero, one, or bounded away from zero and one is desirable. The performance of the threshold linear option is likely due to the combination of a hard threshold with nonuniform choices of the nonzero non-one $q_i$'s.

We remark that a small blocklength $n=20$ is chosen since the current implementation of the encoding and decoding algorithm has exponential time complexity.  Nevertheless, having a good performance for a small $n$ shows that the WPC codes can be applied to practical delay-constrained communication settings where a small blocklength is desirable. The average running time of each trial in our experiment is about $0.14$ second, efficient enough for practical applications. 

\section{Low-density Weighted Parity-Check Codes with Belief Propagation\label{sec:ldpc}}
In the previous section, we have seen the experiments on WPC codes for a small blocklength $n=20$. To apply the WPC codes for a larger blocklength, we require more efficient coding algorithms.
In this section, we describe a sparse construction for the weighted parity-check code similar to the low-density parity-check (LDPC) codes~\cite{gallager1962low,mackay1999good}, which we call \emph{low-density WPC codes}, where
the belief propagation algorithm can be applied to improve the efficiency of the decoding algorithm.

We first briefly review the conventional belief propagation decoding algorithm for LDPC codes \cite{Kschischang2001, Loeliger2004,richardson_urbanke_2008}.
Consider a sparse parity-check matrix $\tilde{\mathbf{H}} \in \mathbb{F}_2^{(n-k) \times n}$.
Given the received sequence $\mathbf{y} = [y_1,\ldots,y_n] \in \mathbb{F}_2^n$, the belief propagation algorithm updates the messages $\lambda_{i,j}$ from the codeword bit $i \in \{1,\ldots,n\}$ to the parity-check bit $j \in \{1,\ldots, n-k\}$, and the messages $\nu_{j,i}$ from the parity-check bit $j$  to the codeword bit $i$, alternately. Initially, $\nu_{j,i}=0$. We then repeat the following two steps (for a fixed number of iterations or until convergence):
\begin{equation}
    \lambda_{i,j} \,\leftarrow\, \log \frac{p_{X|Y}(0|y_i)}{p_{X|Y}(1|y_i)} + \sum_{j' \in \{1,\ldots, n-k\}\backslash\{j\}: \, \tilde{\mathbf{H}}_{j',i} = 1} \nu_{j',i} \label{eq:update_lambda}
\end{equation}
for $(i,j) \in \{1,\ldots,n\} \times \{1,\ldots, n-k\}$ where $\tilde{\mathbf{H}}_{j,i}=1$, and
\begin{equation}
\nu_{j,i} \,\leftarrow\, 2 \cdot \arctanh{\Bigg(\prod_{i' \in \{1,\ldots, n\}\backslash\{i\}: \, \tilde{\mathbf{H}}_{j,i'} = 1}\tanh(\lambda_{i',j}/2)\Bigg)} \label{eq:update_nu}
\end{equation}
for $(i,j) \in \{1,\ldots,n\} \times \{1,\ldots, n-k\}$ where $\tilde{\mathbf{H}}_{j,i}=1$. At the end, we decode to $\hat{x}_i = 0$ if 
\begin{equation}
    \log \frac{p_{X|Y}(0|y_i)}{p_{X|Y}(1|y_i)} + \sum_{j \in \{1,\ldots, n-k\}: \, \tilde{\mathbf{H}}_{j,i} = 1} \nu_{j,i} > 0, \label{eq:bp_decode}
\end{equation}
and $\hat{x}_i = 1$ otherwise. Please refer to \cite{Kschischang2001, Loeliger2004,richardson_urbanke_2008} for a more detailed description of the belief propagation algorithm.

We now combine the belief propagation algorithm with the weighted parity-check codes for channels with state in Section \ref{sec:state}, applied on binary-Hamming information embedding in Section \ref{subsec:info_embed}. We first describe the decoding algorithm, where the decoder observes $\mathbf{y} = [y_1,\ldots,y_n] \in \mathbb{F}_2^n$ and wants to decode $\mathbf{m} \in \mathbb{F}_2^n$. Although $\mathbf{H} \in \mathbb{F}_2^{n \times n}$ is a square matrix for WPC code, the first $k$ bits of $\mathbf{x}\mathbf{H}^T$ are the message bits in $\mathbf{m}$ \eqref{eq:pq_state}, which the decoder does not observe. Therefore, the decoder only has information on the last $n-k$ bits of $\mathbf{x}\mathbf{H}^T$. The actual parity-check matrix according to the decoder is $\tilde{\mathbf{H}} \in \mathbb{F}_2^{(n-k) \times n}$, formed by the last $n-k$ rows of $\mathbf{H}$. The main difference between the WPC codes and the aforementioned belief propagation algorithm is that the parity-check bits here are attached with the weights $\mathbf{q}=[q_1,\ldots,q_{n-k}]$, and the prior distribution of the parity-check bit $(\mathbf{x}\tilde{\mathbf{H}}^T)_j$ is $\mathrm{Bern}(q_j)$ for $j=1,\ldots,n-k$. To incorporate this prior distribution to the belief propagation algorithm, for each parity-check bit $j$, we can create a fictitious codeword bit that is only connected to that parity-check bit, and is observed to be $0$ after passing through a binary symmetric channel with crossover probability $q_j$ (and hence has a log-likelihood ratio $\log((1-q_j)/q_j)$). Adding this fictitious codeword bit to the product \eqref{eq:update_nu}, since $\tanh(\log((1-q_j)/q_j)/2) = 1-2q_j$, we would change \eqref{eq:update_nu} to
\begin{equation}
\nu_{j,i} \,\leftarrow\, 2 \cdot \arctanh{\Bigg((1-2q_j)\prod_{w \in \{1,\ldots, n\}\backslash\{i\}: \, \tilde{\mathbf{H}}_{j,w} = 1}\tanh(\lambda_{w,j}/2)\Bigg)}. \label{eq:update_nu2}
\end{equation}
The update rule \eqref{eq:update_lambda} for $\lambda_{i,j}$ and the decoding rule \eqref{eq:bp_decode} remains the same.

For the encoding function, since the encoder also knows the first $k$ bits of $\mathbf{x}\mathbf{H}^T$ which are given by the message $\mathbf{m}$, it can run a similar belief propagation algorithm as the decoder, with all $n$ parity-check bits (where there are $k$ more parity-check bits given by the message), and use the state sequence $\mathbf{s}$ instead of the channel output $\mathbf{y}$ to calculate the likelihood of $\mathbf{x}$. Alternatively, we may simply compute the query function \eqref{eq:fH} by randomly generating a number of codewords $\mathbf{x}$ using a method described later, and choosing the one that gives the largest $w_{\mathbf{p}_e(\mathbf{m},\mathbf{s})}(\mathbf{x})w_{\mathbf{q}_e(\mathbf{m},\mathbf{s})}(\mathbf{x}\mathbf{H}^{T})$. A suitable way to generate $\mathbf{x}$ would be to first generate the parity-check bits $\mathbf{v} \in \mathbb{F}_2^n$ where $v_i \sim \mathrm{Bern}((\mathbf{q}_e(\mathbf{m},\mathbf{s}))_i)$, and take $\mathbf{x}=\mathbf{v}(\mathbf{H}^{T})^{-1}$. This way, we can ensure that the first $k$ bits of $\mathbf{x}\mathbf{H}^T$ coincide with the message. We observe in the experiments that this random encoding algorithm is more efficient than using belief propagation for encoding.

For the sparse parity-check matrix $\mathbf{H} \in \mathbb{F}_2^{n \times n}$, we can simply generate it by sampling a random regular bipartite graph. We describe a simple algorithm to do so. Fix a degree $d\ge 3$ which must be an odd number. First, we generate permutation matrices $\mathbf{K}_i \in \mathbb{F}_2^{n \times n}$ for $i=1,\ldots,d$ uniformly where the set of positions of ones in $\mathbf{K}_i$ are disjoint. Then, we take $\mathbf{H} = \sum_{i=1}^d \mathbf{K}_i$, which has $nd$ ones. If $\mathbf{H}$ is not full-rank, repeat this process until we obtain a full-rank $\mathbf{H}$ (note that $k$ must be odd for $\mathbf{H}$ to be full-rank). The matrix $\mathbf{H}$ is the adjacency matrix of a regular bipartite graph with degree $d$.

\begin{figure}
    \centering
    \includegraphics[width=1.0\linewidth]{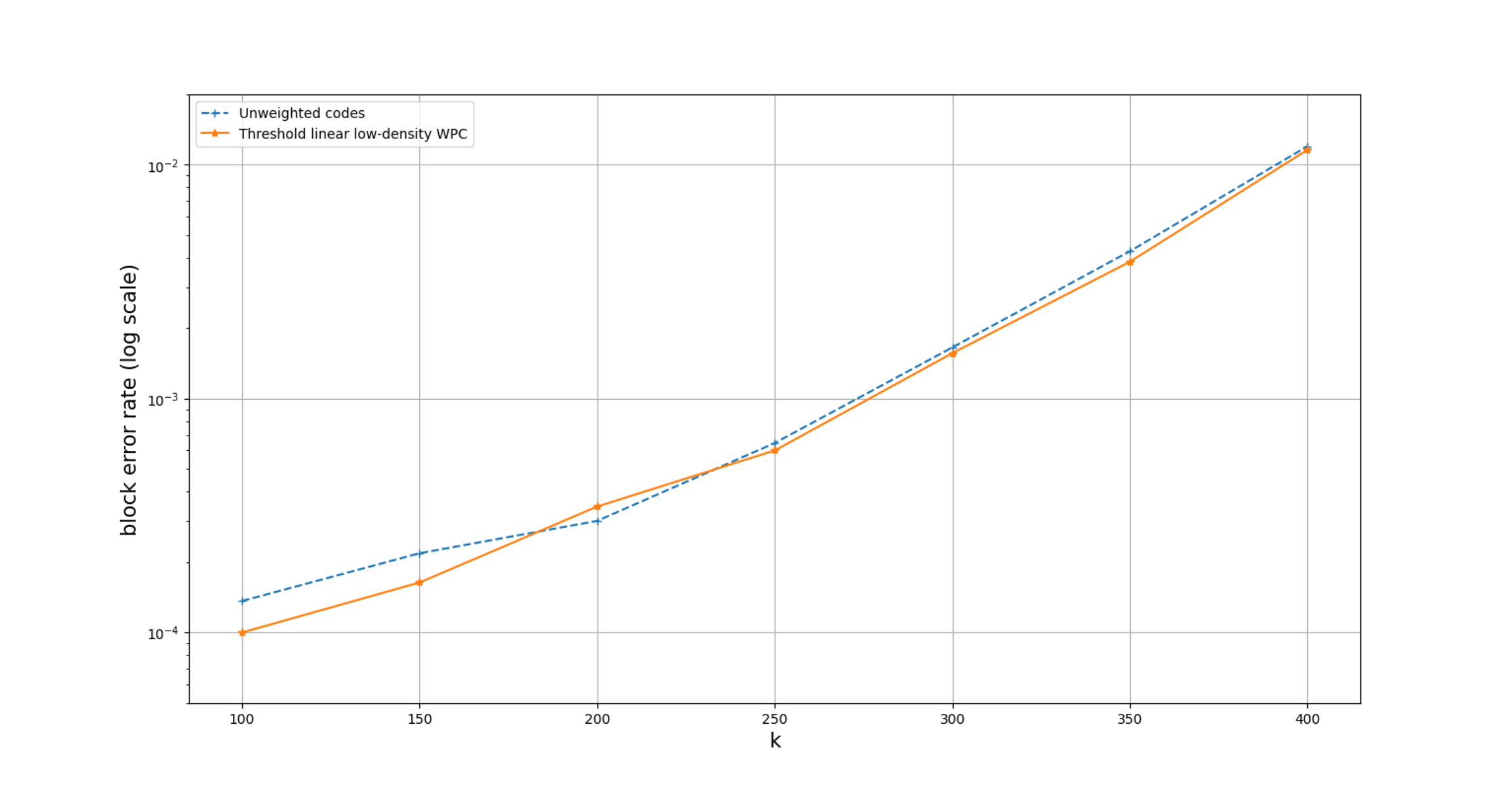}
    \caption{Performance evaluation for the threshold linear low-density weighted parity-check codes compared to the unweighted codes with $n=1000$, $D=0.4$, $\beta=0.05$. 
    Each data point is computed using $5500$ trials.}
    \label{fig::sparseldpc}
    \vspace{-8pt}
\end{figure}

\begin{figure}
    \centering
    \includegraphics[width=1.0\linewidth]{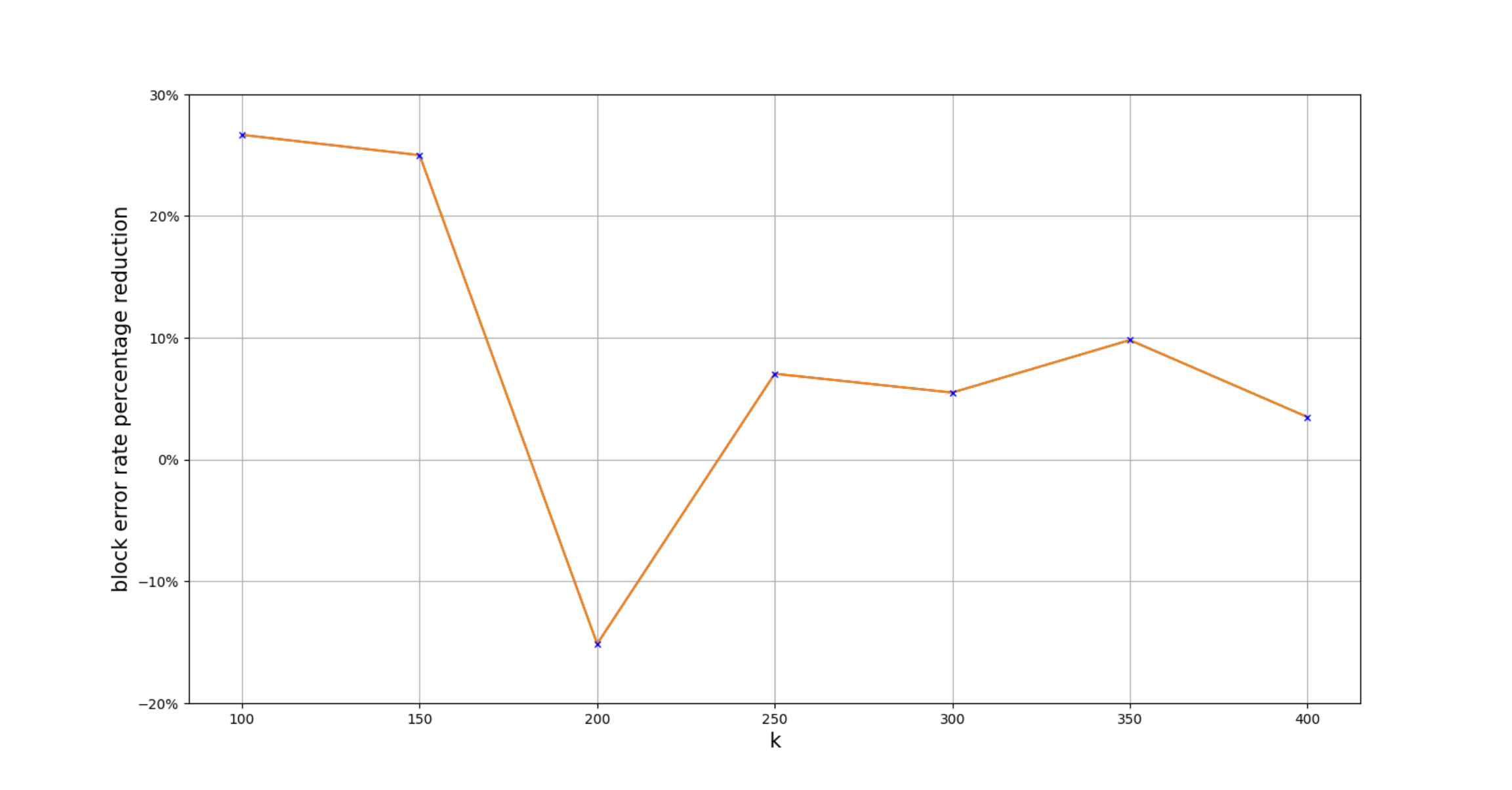}
    \caption{The percentage reduction of the block error rates for the threshold linear low-density weighted parity-check codes compared to the unweighted codes with $n=1000$, $D=0.4$, $\beta=0.05$. 
    Each data point is computed using $5500$ trials.}
    \label{fig::sparseldpcPerc}
    \vspace{-8pt}
\end{figure}

We perform experiments on binary-Hamming information embedding in Section \ref{subsec:info_embed}, where the blocklength is $n=1000$, the message lengths are $k=100,150,200,\ldots,400$, the degree is $d=11$, the crossover probability of the channel is $\beta = 0.05$, and the target cost per symbol is $D=0.4$. We compare the threshold linear low-density WPC codes \eqref{eq:thres_linear} to the unweighted method (i.e., the threshold method \eqref{eq:thres_gamma}), which corresponds to the nested linear code \cite{Zamir2002Nested,barron2003duality} applied on a sparse matrix where there are no weighted parity-check bits (all parity-check bits are fixed to $0$ or $1$). We apply the random encoding algorithm and the belief propagation decoding algorithm.\footnote{In the random encoding algorithm, $8000$ samples of $\mathbf{x}$ are generated. To improve the decoding accuracy, the belief propagation decoding algorithm is run $5$ times, each time with randomly initialized $\nu_{j,i}$.} The experiment was implemented in Python using a modified version of the pyldpc package for LDPC codes \cite{Janati2020}. We perform $5500$ trials per data point to obtain the average block error rates. 

Figure~\ref{fig::sparseldpc} shows the block error rates of the two methods for different message lengths $k$. Figure~\ref{fig::sparseldpcPerc} shows the percentage reduction of the block error rate of the threshold linear low-density WPC code compared to the unweighted method. We can see that the threshold linear low-density WPC codes outperform the unweighted method except for $k=200$. The improvement is especially significant for smaller $k$, where the threshold linear low-density WPC codes reduce the error rate by $26.7\%$ for $k=100$, and  $25.0\%$ for $k=150$, compared to the unweighted method. While the improvement might not appear to be large, it is still perhaps surprising that the error rate can be reduced via a small change to the belief propagation algorithm by attaching weights to the parity-check bits. Note that the average cost for each $k$ is measured to be within the range $400.05 \pm0.05$, which is very close to the expected cost $nD=400$, so we do not plot the cost here.

\ifshortver
\else

\section{Wyner-Ziv Problem}

Consider the Wyner-Ziv problem~\cite{wyner1976ratedistort}, where 
there is a 2-discrete memoryless source $(\mathbf{x},\mathbf{y})$ where $(x_i,y_i) \in \mathcal{X}\times\mathcal{Y}$, $(x_i,y_i)\stackrel{iid}{\sim} P_{X,Y}$ for $i=1,\ldots,n$.
The encoder observes $\mathbf{x}$ and produces the message $\mathbf{m}\in\mathbb{F}_{2}^{k}$.
The decoder observes $\mathbf{m}$ and the side information $\mathbf{y}$, and recovers $\hat{\mathbf{z}}\in \mathcal{Z}^n$, subject to a distortion constraint
$\mathbf{E}[\sum_{i=1}^n d(x_i,\hat{z}_i)] \le nD$, where $d : \mathcal{X}\times \mathcal{Z} \to [0,\infty)$.
The goal
is to design a coding scheme satisfying the distortion constraint as $n\to\infty$ when the message length is $k=\lfloor nR\rfloor$,
where $R>0$ is the rate.

Assume $\mathcal{Z} = \mathbb{F}_2$. Consider the following coding scheme. 
Recall that the full parity-check matrix $\mathbf{H}$ is a uniformly
chosen random full-rank matrix.
The encoder uses the encoder codeword and parity bias
functions $\mathbf{p}_{e}(\mathbf{x})$, $\mathbf{q}_{e}(\mathbf{x})$
to obtain $\mathbf{z}=f_{\mathbf{H}}(\mathbf{p}_{e}(\mathbf{x}),\,\mathbf{q}_{e}(\mathbf{x}))$, $\mathbf{m}=[(\mathbf{z}\mathbf{H}^{T})_{1},\ldots,\,(\mathbf{z}\mathbf{H}^{T})_{k}]$. The decoder uses the decoder
codeword and parity bias functions $\mathbf{p}_{d}(\mathbf{y},\mathbf{m})$,
$\mathbf{q}_{d}(\mathbf{y},\mathbf{m})$ to obtain $\hat{\mathbf{z}}=f_{\mathbf{H}}(\mathbf{p}_{d}(\mathbf{y},\mathbf{m}),\,\mathbf{q}_{d}(\mathbf{y},\mathbf{m}))$.
We take
\begin{align*}
\mathbf{p}_{e}(\mathbf{x}) & =[p_{e}(x_{1}),\ldots,p_{e}(x_{n})],\\
\mathbf{q}_{e}(\mathbf{x}) & =[\frac{1}{2}\mathbf{1}^{k},\,\mathbf{q}],\\
\mathbf{p}_{d}(\mathbf{y},\mathbf{m}) & =[p_{d}(y_{1}),\ldots,p_{d}(y_{n})],\\
\mathbf{q}_{d}(\mathbf{y},\mathbf{m}) & =[\mathbf{m},\,\mathbf{q}],
\end{align*}
where $p_{e}:\mathcal{X}\to[0,1]$, $p_{d}:\mathcal{Y}\to[0,1]$ are
parameters of the encoder and decoder, and $\mathbf{q}=[q_{1},\ldots,q_{n-k}]$,
where $q_{i}\sim P_{Q}$ i.i.d., 
and $P_{Q}$ is the parity
bias distribution, a distribution over $[0,1]$ symmetric about $1/2$.

The following theorem shows that the code can achieve the optimal rate in~\cite{wyner1976ratedistort}.
\begin{thm}
\label{thm:capacity_wz}Assume $|\mathcal{X}|,|\mathcal{Y}|<\infty$, $\mathcal{Z} = \mathbb{F}_2$. Fix any $P_{Z|X}$. Consider the weighted parity-check code, where  $p_{e}(x)=P_{Z|X}(1|x)$, $p_{d}(y)=P_{Z|Y}(1|y)$, and $P_{Q}$ is a discrete distribution over $[0,1]$ symmetric about $1/2$ with finite support satisfying
\begin{equation}
\mathbf{E}[H_{b}(Q)]=1-\frac{H(Z|X)}{1-R}.\label{eq:wz_ehbq}
\end{equation}
For any $R>I(X;Z|Y)$, as $n\to\infty$, the empirical joint distribution of $\{(x_{i},\hat{z}_{i})\}_{i=1,\ldots,n}$
tends to $P_{X}P_{Z|X}$ in probability.
\end{thm}

To prove Theorem~\ref{thm:capacity_wz}, we use the following lemma which gives a sufficient condition for the probability of error to tend to $0$. 
\begin{lem}
\label{lem:wynerziv}Consider the weighted parity-check code for the Wyner-Ziv problem,
where $|\mathcal{X}|,|\mathcal{Y}|<\infty$, $\mathcal{Z} = \mathbb{F}_2$, and $P_{Q}$ is a discrete distribution over $[0,1]$ symmetric about $1/2$ with finite support. Let
$(X,Y)\sim P_{X,Y}$, $Z|X\sim P_{Z|X}$, $Q\sim P_{Q}$, $V\in \{0,1\}$, $V|Q\sim P_{V|Q}$, where $(P_{Z|X},P_{V|Q})$ is the 
minimizer of 
\begin{align}
 & \mathbf{E}\left[H_{b}(Z,p_{e}(X))\right]+(1-R)\mathbf{E}\left[H_{b}(V,Q)\right],\label{eq:wz_ehb_min}
\end{align}
subject to
\begin{equation}
H(Z|X)+(1-R)H(V|Q)\ge 1 - R.\label{eq:wz_ehb_min_cons}
\end{equation}
If the minimizer of~\eqref{eq:wz_ehb_min} is unique, and for all $P_{\tilde{Z}|Y}$, $P_{\tilde{V}|Q}$ 
satisfying
\begin{equation}
H(\tilde{Z}|Y)+(1-R)H(\tilde{V}|Q)\ge 1,\label{eq:wz_hzy_hvq}
\end{equation}
we have
\begin{align}
 & \mathbf{E}[H_{b}(\tilde{Z},p_{d}(Y))]+(1-R)\mathbf{E}[H_{b}(\tilde{V},Q)]\nonumber\\
 & >\mathbf{E}[H_{b}(Z,p_{d}(Y))]+(1-R)\mathbf{E}[H_{b}(V,Q)],\label{eq:wz_dec_ineq}
\end{align}
then as $n\to\infty$, the empirical joint distribution of $\{(x_{i},\hat{z}_{i})\}_{i=1,\ldots,n}$
tends to $P_{X}P_{Z|X}$ in probability.
\end{lem}

\begin{IEEEproof}
The proof is similar to the proof of Lemma \ref{lem:state}. We
assume that the encoder and the decoder use \eqref{eq:fxHb} instead
of \eqref{eq:fH}. Let $\mathbf{z}$ be the sequence chosen by the
encoder, and $[\mathbf{m},\mathbf{v}]=\mathbf{z}\mathbf{H}^{T}\oplus\mathbf{b}$.
Using the same arguments as Lemma \ref{lem:state}, we can assume
$(\mathbf{x},\mathbf{y},\mathbf{z})\in\mathcal{T}_{\epsilon_{1}}^{(n)}(P_{X,Y,Z})$,
$(\mathbf{q},\mathbf{v})\in\mathcal{T}_{\epsilon_{1}}^{((1-R)n)}(P_{Q,V})$
are jointly typical for some $\epsilon_{1}>0$.

It is left to show that $\hat{\mathbf{z}}=\mathbf{z}$ with high probability.
Unfortunately, unlike \eqref{eq:enc_dep}, now the encoder depends on the
whole $(\mathbf{H},\mathbf{b})$, which results in dependency between
the encoder and the decoder. Hence we use another approach to bound
the decoding error rate. For a set $\mathcal{A}$, write $\mathcal{A}_{\neq}^{2}:=\{(a_{1},a_{2})\in\mathcal{A}^{2}:\,a_{1}\neq a_{2}\}$.
Let $\mathbf{m}_{0}\sim\mathrm{Unif}(\mathbb{F}_{2}^{Rn})$ be independent
of $(\mathbf{v}_{1},\mathbf{v}_{2})\sim\mathrm{Unif}((\mathbb{F}_{2}^{(1-R)n})_{\neq}^{2})$.
Let $\mathbf{z}_{1}:=([\mathbf{m}_{0},\mathbf{v}_{1}]\oplus\mathbf{b})\mathbf{H}^{-T}$,
$\mathbf{z}_{2}:=([\mathbf{m}_{0},\mathbf{v}_{2}]\oplus\mathbf{b})\mathbf{H}^{-T}$.
We have $(\mathbf{z}_{1},\mathbf{z}_{2})\sim\mathrm{Unif}((\mathbb{F}_{2}^{n})_{\neq}^{2})$
independent of $(\mathbf{m}_{0},\mathbf{v}_{1},\mathbf{v}_{2})$ due
to the randomness in $\mathbf{H},\mathbf{b}$. By asymptotic equipartition
property,
\begin{equation}
\mathbf{P}\left((\mathbf{x},\mathbf{z}_{1})\in\mathcal{T}_{\epsilon_{1}}^{(n)}(P_{X,Z})\right)\le2^{-n(1-H(Z|X)-\delta(\epsilon_{1}))},\label{eq:xz1_in_T}
\end{equation}
\begin{equation}
\mathbf{P}\left((\mathbf{q},\mathbf{v}_{1})\in\mathcal{T}_{\epsilon_{1}}^{((1-R)n)}(P_{Q,V})\right)\le2^{-(1-R)n(1-H(V|Q)-\delta(\epsilon_{1}))}.\label{eq:qv1_in_T}
\end{equation}
Let $(\hat{Y},\hat{Z}_{2})\sim\hat{P}_{\mathbf{y},\mathbf{z}_{2}}$,
$(\hat{Q},\hat{V}_{2})\sim\hat{P}_{\mathbf{q},\mathbf{v}_{2}}$. By
Sanov's theorem,
\begin{align}
 & \mathbf{P}\big(H(\hat{Z}_{2}|\hat{Y})+(1-R)H(\hat{V}_{2}|\hat{Q})\le1-\epsilon_{2}\big)\nonumber \\
 & =\mathbf{P}\big(D(P_{\hat{Z}_{2}|\hat{Y}}\Vert\mathrm{Bern}(1/2)|\hat{Y})\nonumber \\
 & \;\;\;\;+(1-R)D(P_{\hat{V}_{2}|\hat{Q}}\Vert\mathrm{Bern}(1/2)|\hat{Q})\ge1-R+\epsilon_{2}\big)\nonumber \\
 & \le2^{-n(1-R+\epsilon_{2}-\delta(\epsilon_{1}))+o(n)}.\label{eq:z2y_v2q_ent}
\end{align}
Note that $\mathbf{v}_{1},\mathbf{v}_{2},\mathbf{z}_{1},\mathbf{z}_{2}$
are almost mutually independent, that is, the probability ratio 
\begin{align*}
 & \frac{dP_{\mathbf{v}_{1},\mathbf{v}_{2},\mathbf{z}_{1},\mathbf{z}_{2}}}{d(P_{\mathbf{v}_{1}}\times P_{\mathbf{v}_{2}}\times P_{\mathbf{z}_{1}}\times P_{\mathbf{z}_{2}})}\\
 & \le\frac{2^{(1-R)n}}{2^{(1-R)n}-1}\cdot\frac{2^{n}}{2^{n}-1}\\
 & \le1+4\cdot2^{-(1-R)n}
\end{align*}
for $(1-R)n\ge10$. Multiplying this with \eqref{eq:xz1_in_T}, \eqref{eq:qv1_in_T}
and \eqref{eq:z2y_v2q_ent}, we have
\begin{align*}
 & \mathbf{P}\big((\mathbf{x},\mathbf{z}_{1})\in\mathcal{T}_{\epsilon_{1}}^{(n)}(P_{X,Z}),\,\\
 & \;\;\;\;\;(\mathbf{q},\mathbf{v}_{1})\in\mathcal{T}_{\epsilon_{1}}^{((1-R)n)}(P_{Q,V}),\\
 & \;\;\;\;\;H(\hat{Z}_{2}|\hat{Y})+(1-R)H(\hat{V}_{2}|\hat{Q})\le1-\epsilon_{2}\big)\\
 & \le(1+4\cdot2^{-(1-R)n})\\
 & \;\;\;\;\cdot2^{-n(3-H(Z|X)-(1-R)H(V|Q)-2R+\epsilon_{2}-3\delta(\epsilon_{1}))+o(n)}\\
 & =(1+4\cdot2^{-(1-R)n})2^{-n(2-R+\epsilon_{2}-3\delta(\epsilon_{1}))+o(n)},
\end{align*}
where the last line is because equality in~\eqref{eq:wz_ehb_min_cons} must hold (or else the
minimizer of \eqref{eq:wz_ehb_min} is not unique). By union bound,
\begin{align}
 & \mathbf{P}\big(\exists\mathbf{m}_{0}\in\mathbb{F}_{2}^{Rn},\,(\mathbf{v}_{1},\mathbf{v}_{2})\in(\mathbb{F}_{2}^{(1-R)n})_{\neq}^{2}.\nonumber \\
 & \;\;\;\;\;(\mathbf{x},\mathbf{z}_{1})\in\mathcal{T}_{\epsilon_{1}}^{(n)}(P_{X,Z}),\,\nonumber \\
 & \;\;\;\;\;(\mathbf{q},\mathbf{v}_{1})\in\mathcal{T}_{\epsilon_{1}}^{((1-R)n)}(P_{Q,V}),\nonumber \\
 & \;\;\;\;\;H(\hat{Z}_{2}|\hat{Y})+(1-R)H(\hat{V}_{2}|\hat{Q})\le1-\epsilon_{2}\big)\label{eq:m0v1v2}\\
 & \le(1+4\cdot2^{-(1-R)n})2^{-n(\epsilon_{2}-3\delta(\epsilon_{1}))+o(n)}.\nonumber 
\end{align}
Hence, by choosing $\epsilon_{1}$ small enough compared to $\epsilon_2$, we can assume there
does not exist $\mathbf{m}_{0},\mathbf{v}_{1},\mathbf{v}_{2}$ satisfying
the conditions in \eqref{eq:m0v1v2} (which happens with probability
approaching $1$). Since $(\mathbf{x},\mathbf{z})\in\mathcal{T}_{\epsilon_{1}}^{(n)}(P_{X,Z})$,
$(\mathbf{q},\mathbf{v})\in\mathcal{T}_{\epsilon_{1}}^{((1-R)n)}(P_{Q,V})$,
we know that $H(\hat{Z}_{2}|\hat{Y})+(1-R)H(\hat{V}_{2}|\hat{Q})>1-\epsilon_{2}$
for any $\mathbf{v}_{2}\neq\mathbf{v}$ (note that $\mathbf{z}_{2}=([\mathbf{m},\mathbf{v}_{2}]\oplus\mathbf{b})\mathbf{H}^{-T}$).
Using the same arguments as Lemma \ref{lem:state}, we can show
that this, together with \eqref{eq:wz_dec_ineq}, implies
\begin{align*}
 & \mathbf{E}\big[H_{b}(\hat{Z}_{2},p_{d}(\hat{Y}))\big]+(1-R)\mathbf{E}\big[H_{b}(\hat{V}_{2},\hat{Q})\big]\\
 & >\mathbf{E}\big[H_{b}(X,p_{d}(Y))\big]+(1-R)\mathbf{E}\big[H_{b}(V,Q)\big]+\epsilon_{3}
\end{align*}
for any $\mathbf{v}_{2}\neq\mathbf{v}$. Hence, the decoder will not
choose $\mathbf{z}_{2}=([\mathbf{m},\mathbf{v}_{2}]\oplus\mathbf{b})\mathbf{H}^{-T}$
over $\mathbf{z}$.
\end{IEEEproof}
\medskip

We now prove Theorem~\ref{thm:capacity_wz}
\begin{IEEEproof}[Proof of Theorem~\ref{thm:capacity_wz}]
Note that $P_{Z|X}(1|x)=p_{e}(x)$, $P_{V|Q}(1|q)=q$ is the unique maximizer of $H(Z|X)-\mathbf{E}[H_{b}(Z,p_{e}(X))]+(1-R)(H(V|Q)-\mathbf{E}[H_{b}(V,Q)])$. By~\eqref{eq:wz_ehbq}, we can deduce that $(P_{Z|X},P_{V|Q})$ is the unique minimizer of \eqref{eq:wz_ehb_min}.
It remains to check~\eqref{eq:wz_dec_ineq}.
We have $\mathbf{E}[H_{b}(Q)]=H(V|Q)$, $\mathbf{E}[H_{b}(Z,p_{d}(Y))]=H(Z|Y)$, and $\mathbf{E}[H_{b}(\tilde{Z},p_{d}(Y))]\ge H(\tilde{Z}|Y)$.
If $R>H(Z|Y)-H(Z|X)$, then
\begin{align*}
 & \mathbf{E}[H_{b}(\tilde{Z},p_{d}(Y))]+(1-R)\mathbf{E}[H_{b}(\tilde{V},Q)]\\
 & \ge H(\tilde{Z}|Y)+(1-R)\mathbf{E}[H_{b}(\tilde{V},Q)]\\
 & \stackrel{(a)}{\ge}1+(1-R)\big(\mathbf{E}[H_{b}(\tilde{V},Q)] - H(\tilde{V}|Q)\big)\\
 & >1-R + H(Z|Y) - H(Z|X) \\
 & \;\;\;\; +(1  - R)\big(\mathbf{E}[H_{b}(\tilde{V},Q)] - H(\tilde{V}|Q)\big)\\
 & \stackrel{(b)}{=}(1-R)H(V|Q)+H(Z|Y) \\
 &\;\;\;\;+(1-R)\big(\mathbf{E}[H_{b}(\tilde{V},Q)] - H(\tilde{V}|Q)\big)\\
 & =\mathbf{E}[H_{b}(Z,p_{d}(Y))]\\
 &\;\;\;\;+(1-R)\big(H(V|Q)+\mathbf{E}[H_{b}(\tilde{V},Q)]-H(\tilde{V}|Q)\big)\\
 & \stackrel{(c)}{\ge}\mathbf{E}[H_{b}(Z,p_{d}(Y))]+(1-R)\mathbf{E}[H_{b}(V,Q)],
\end{align*}
where (a) is by \eqref{eq:wz_hzy_hvq}, (b) is by~\eqref{eq:wz_ehbq}, and (c) is because $P_{V|Q}(1|q)=q$
maximizes $H(V|Q)-\mathbf{E}[H_{b}(V,Q)]$. The result follows from
Lemma \ref{lem:wynerziv}.
\end{IEEEproof}

\fi

\section{Conclusion and Discussions}

In this paper, we introduced the weighted parity-check (WPC) codes, where weights are attached to the parity-check bits. It is applicable to channels with state~\cite{gelfand1980coding}, asymmetric channels, and the Wyner-Ziv problem~\cite{wyner1976ratedistort}. We proved that the WPC codes are capacity-achieving for (symmetric or asymmetric) channels with state and the Wyner-Ziv problem. We performed experiments which show the reduction of error rate of the proposed WPC codes compared to the nested linear codes \cite{Zamir2002Nested,barron2003duality} for channels with state. We also performed experiments on a sparse construction for the WPC codes where the belief propagation algorithm can be applied.

A potential future direction is to apply the WPC codes to multiuser channels, such as broadcast channels~\cite{cover1972broadcast} and interference channels~\cite{ahlswede1974capacity}. Another direction is to perform a theoretical analysis on the WPC codes with sparse parity-check matrices (recall that the theorems in this paper assume that the parity-check matrix is a random dense matrix). The key is to prove a version of Lemma~\ref{lem:ab_bd} that applies to sparse matrices, which does not appear to be straightforward.

\section{Acknowledgement}

The work of Cheuk Ting Li was supported in part by the Hong Kong Research Grant Council Grant ECS No. CUHK 24205621, and the Direct Grant for Research, The Chinese University of Hong Kong (Project ID: 4055133).
The authors would like to thank the anonymous reviewers of the short conference version of this paper~\cite{ling2022weighted} for their valuable suggestions.

\medskip{}

\ifshortver
\else
\appendix

\subsection{Proof of Lemma \ref{lem:ab_bd} \label{subsec:pf_ab_bd}}

Define a sequence of vector subspaces $\{\mathbf{0}\}=V_{0}\subseteq V_{1}\subseteq\cdots\subseteq V_{n}=\mathbb{F}_{l}^{n}$
recursively by taking $V_{i}$ ($i=1,\ldots,n$) to be the subspace
satisfying $V_{i}\supseteq V_{i-1}$, $\mathrm{dim}(V_{i})=i$, and
maximizes $|B\cap V_{i}|$. Note that if we choose a subspace $\tilde{V}_{i}$
satisfying $V_{i-1}\subseteq\tilde{V}_{i}\subseteq V_{i+1}$ uniformly
at random, then for any $\mathbf{x}\in V_{i+1}\backslash V_{i-1}$,
\[
\mathbf{P}(\mathbf{x}\in\tilde{V}_{i})=\frac{l^{i}-l^{i-1}}{l^{i+1}-l^{i-1}}=\frac{1}{l+1}.
\]
Hence 
\[
\mathbf{E}\left[\left|B\cap(\tilde{V}_{i}\backslash V_{i-1})\right|\right]=\frac{1}{l+1}\left|B\cap(V_{i+1}\backslash V_{i-1})\right|,
\]
\[
\mathbf{E}\left[\left|B\cap\tilde{V}_{i}\right|\right]=\frac{1}{l+1}\left|B\cap V_{i+1}\right|+\frac{l}{l+1}\left|B\cap V_{i-1}\right|.
\]
By the maximality of $V_{i}$, 
\[
\left|B\cap V_{i}\right|\ge\frac{1}{l+1}\left|B\cap V_{i+1}\right|+\frac{l}{l+1}\left|B\cap V_{i-1}\right|,
\]
\[
\left|B\cap(V_{i+1}\backslash V_{i})\right|\le l\left|B\cap(V_{i}\backslash V_{i-1})\right|.
\]
Let 
\[
\tilde{B}_{i}:=B\cap(V_{i}\backslash V_{i-1}).
\]
We have $|\tilde{B}_{i+1}|\le l|\tilde{B}_{i}|$. Assume $i^{*}\in\{1,\ldots,n\}$
attains the maximum of $|\tilde{B}_{i}|$. We have $|\tilde{B}_{i^{*}}|\ge|B|/n$.
Since $|\tilde{B}_{i+1}|\le l|\tilde{B}_{i}|$ and $|\tilde{B}_{1}|=1$,
we have 
\begin{equation}
|\{i:|\tilde{B}_{i}|\ge\xi\}|\ge\log_{l}\frac{|\tilde{B}_{i^{*}}|}{\xi}\ge\log_{l}\frac{|B|}{n\xi}\label{eq:bt_sz_bd}
\end{equation}
for any $1\le\xi\le|B|/n$.

Note that for any $\mathbf{x}\in\tilde{B}_{i}$, conditional on $\{\mathbf{H}\mathbf{y}\}_{\mathbf{y}\in V_{i-1}}$,
we have $\mathbf{H}\mathbf{x}$ uniformly distributed over $\mathbb{F}_{l}^{n}\backslash\mathbf{H}V_{i-1}$.
Hence, for any $\mathbf{x}\in\tilde{B}_{i}$, 
\begin{align*}
 & \mathbf{P}\left(\mathbf{H}\mathbf{x}\in A\,\big|\,A\cap\mathbf{H}\left(B\cap V_{i-1}\right)\neq\emptyset\right)\\
 & =\mathbf{E}\left[\mathbf{P}\left(\mathbf{H}\mathbf{x}\in A\,\big|\,\{\mathbf{H}\mathbf{y}\}_{\mathbf{y}\in V_{i-1}}\right)\,\bigg|\,A\cap\mathbf{H}\left(B\cap V_{i-1}\right)\neq\emptyset\right]\\
 & =\mathbf{E}\left[\frac{\left|A\backslash\mathbf{H}V_{i-1}\right|}{l^{n}-l^{i-1}}\,\bigg|\,A\cap\mathbf{H}\left(B\cap V_{i-1}\right)\neq\emptyset\right]\\
 & \le\frac{|A|}{l^{n}-l^{i-1}}.
\end{align*}
Let 
\[
\rho_{i}=\mathbf{P}\left(A\cap\mathbf{H}\left(B\cap V_{i}\right)\neq\emptyset\right),
\]
and $\tilde{B}_{i}=\{\mathbf{x}_{1},\ldots,\mathbf{x}_{|\tilde{B}_{i}|}\}$.
For any $m\le|\tilde{B}_{i}|$, we have 
\begin{align*}
\rho_{i} & =\rho_{i-1}+\mathbf{P}\left(A\cap\mathbf{H}\left(B\cap V_{i}\right)\neq\emptyset,\,A\cap\mathbf{H}\left(B\cap V_{i-1}\right)=\emptyset\right)\\
 & =\rho_{i-1}+\sum_{j=1}^{|\tilde{B}_{i}|}\mathbf{P}\left(\mathbf{H}\mathbf{x}_{j}\in A,\,A\cap\mathbf{H}\left(B\cap V_{i-1}\right)=\emptyset,\,\forall j'<j.\mathbf{H}\mathbf{x}_{j'}\notin A\right)\\
 & \ge\rho_{i-1}+\sum_{j=1}^{m}\left(\mathbf{P}\left(\mathbf{H}\mathbf{x}_{j}\in A,\,A\cap\mathbf{H}\left(B\cap V_{i-1}\right)=\emptyset\right)-\sum_{j'=1}^{j-1}\mathbf{P}\left(\mathbf{H}\mathbf{x}_{j}\in A,\,\mathbf{H}\mathbf{x}_{j'}\in A\right)\right)\\
 & =\rho_{i-1}+\sum_{j=1}^{m}\left(\frac{|A|}{l^{n}-1}-\mathbf{P}\left(\mathbf{H}\mathbf{x}_{j}\in A,\,A\cap\mathbf{H}\left(B\cap V_{i-1}\right)\neq\emptyset\right)-(j-1)\frac{|A|}{l^{n}-1}\cdot\frac{|A|-1}{l^{n}-2}\right)\\
 & \ge\rho_{i-1}+\sum_{j=1}^{m}\left(\frac{|A|}{l^{n}-1}-\rho_{i-1}\frac{|A|}{l^{n}-l^{i-1}}-(j-1)\left(\frac{|A|}{l^{n}-1}\right)^{2}\right)\\
 & \ge\rho_{i-1}\left(1-\frac{m|A|}{l^{n}-l^{i-1}}\right)+\frac{m|A|}{l^{n}-1}-\frac{m(m-1)}{2}\left(\frac{|A|}{l^{n}-1}\right)^{2}\\
 & =\rho_{i-1}\left(1-\frac{m|A|}{l^{n}-l^{i-1}}\right)+\frac{m|A|}{l^{n}-1}\left(1-\frac{(m-1)|A|}{2(l^{n}-1)}\right).
\end{align*}
If 
\[
|\tilde{B}_{i}|\ge\gamma\frac{l^{n}-1}{|A|}
\]
for some $0<\gamma<1$, substituting 
\[
m=\left\lceil \gamma\frac{l^{n}-1}{|A|}\right\rceil ,
\]
we have 
\begin{align*}
\rho_{i} & \ge\rho_{i-1}\left(1-\left\lceil \gamma\frac{l^{n}-1}{|A|}\right\rceil \frac{|A|}{l^{n}-l^{i-1}}\right)+\left\lceil \gamma\frac{l^{n}-1}{|A|}\right\rceil \frac{|A|}{l^{n}-1}\left(1-\gamma\frac{l^{n}-1}{|A|}\frac{|A|}{2(l^{n}-1)}\right)\\
 & =\rho_{i-1}\left(1-\left\lceil \gamma\frac{l^{n}-1}{|A|}\right\rceil \frac{|A|}{l^{n}-l^{i-1}}\right)+\left\lceil \gamma\frac{l^{n}-1}{|A|}\right\rceil \frac{|A|}{l^{n}-1}\left(1-\frac{\gamma}{2}\right).
\end{align*}
Substituting 
\[
m=\left\lfloor \gamma\frac{l^{n}-1}{|A|}\right\rfloor ,
\]
we have 
\[
\rho_{i}\ge\rho_{i-1}\left(1-\left\lfloor \gamma\frac{l^{n}-1}{|A|}\right\rfloor \frac{|A|}{l^{n}-l^{i-1}}\right)+\left\lfloor \gamma\frac{l^{n}-1}{|A|}\right\rfloor \frac{|A|}{l^{n}-1}\left(1-\frac{\gamma}{2}\right).
\]
Taking weighted average of these two bounds on $\rho_{i}$, we have
\begin{align}
\rho_{i} & \ge\rho_{i-1}\left(1-\gamma\frac{l^{n}-1}{|A|}\frac{|A|}{l^{n}-l^{i-1}}\right)+\gamma\frac{l^{n}-1}{|A|}\frac{|A|}{l^{n}-1}\left(1-\frac{\gamma}{2}\right)\nonumber \\
 & =\rho_{i-1}\left(1-\gamma\frac{l^{n}-1}{l^{n}-l^{i-1}}\right)+\gamma\left(1-\frac{\gamma}{2}\right)\nonumber \\
 & \ge\rho_{i-1}\left(1-\frac{\gamma}{1-l^{i-1-n}}\right)+\gamma\left(1-\frac{\gamma}{2}\right)\nonumber \\
 & \ge\rho_{i-1}\left(1-\frac{\gamma}{1-l^{-c-1}}\right)+\gamma\left(1-\frac{\gamma}{2}\right)\label{eq:rho_recur}
\end{align}
if $i\le n-c$, where $c\ge0$. Let 
\[
L:=\left|\left\{ i\in\{1,\ldots,\lfloor n-c\rfloor\}:\,|\tilde{B}_{i}|\ge\gamma\frac{l^{n}}{|A|}\right\} \right|.
\]
Note that $\rho_{i}$ is a nondecreasing sequence. By applying \eqref{eq:rho_recur}
$L$ times (for each $i$ in the above set), we have 
\begin{align*}
\rho_{n} & \ge\gamma\left(1-\frac{\gamma}{2}\right)\sum_{j=0}^{L-1}\left(1-\frac{\gamma}{1-l^{-c-1}}\right)^{j}\\
 & =\gamma\left(1-\frac{\gamma}{2}\right)\frac{1-\left(1-\frac{\gamma}{1-l^{-c-1}}\right)^{L}}{\frac{\gamma}{1-l^{-c-1}}}\\
 & =\left(1-\frac{\gamma}{2}\right)\left(1-l^{-c-1}\right)\left(1-\left(1-\frac{\gamma}{1-l^{-c-1}}\right)^{L}\right)\\
 & \ge\left(1-\frac{\gamma}{2}\right)\left(1-l^{-c-1}\right)\left(1-\left(1-\gamma\right)^{L}\right).
\end{align*}
By \eqref{eq:bt_sz_bd}, if 
\begin{equation}
1\le\gamma\frac{l^{n}}{|A|}\le\frac{|B|}{n},\label{eq:gamma_bd}
\end{equation}
then we have 
\begin{align*}
L & \ge\log_{l}\frac{|A||B|}{nl^{n}\gamma}-c-1\\
 & =\log_{l}\frac{|A||B|}{n\gamma}-n-c-1.
\end{align*}
Taking $\gamma=l^{-c}$, 
\begin{align*}
\rho_{n} & \ge\left(1-l^{-c}/2\right)^{2}\left(1-\left(1-l^{-c}\right)^{\log_{l}\frac{|A||B|}{nl^{-c}}-n-c-1}\right)\\
 & \ge\left(1-l^{-c}\right)\left(1-\left(1-l^{-c}\right)^{\theta}\right)\\
 & =\left(1-l^{-c}\right)\left(1-\exp\left(\theta\ln\left(1-l^{-c}\right)\right)\right)\\
 & \ge\left(1-l^{-c}\right)\left(1-\exp\left(-\theta l^{-c}\right)\right)
\end{align*}
as long as $\theta\ge0$, where 
\[
\theta:=\log_{l}\frac{|A||B|}{n}-n-1.
\]
Taking 
\[
c=\log_{l}\theta-\log_{l}\ln\theta,
\]
we have 
\begin{align*}
\rho_{n} & \ge\left(1-\frac{\ln\theta}{\theta}\right)\left(1-\exp\left(-\ln\theta\right)\right)\\
 & \ge1-\frac{1+\ln\theta}{\theta}.
\end{align*}
The requirement $c\ge0$ is always satisfied as long as $\theta>1$.
The requirement \eqref{eq:gamma_bd} becomes to 
\[
1\le\frac{\ln\theta}{\theta}\frac{l^{n}}{|A|}\le\frac{|B|}{n},
\]
which can be written as 
\[
\frac{\ln\theta}{\theta}\ge\frac{|A|}{l^{n}}
\]
and 
\[
\frac{\ln\theta}{\theta}\le\frac{|A||B|}{nl^{n}}=l^{\theta+1},
\]
which is always satisfied as long as $\theta\ge1$. The result follows.

\medskip{}
\fi

\bibliographystyle{IEEEtran}
\bibliography{ref}

\end{document}